\documentclass[a4paper,11pt]{article}
\pdfoutput=1 

\usepackage{jcappub} 

\usepackage{scalerel}
\usepackage[T1]{fontenc} 
\usepackage{amsfonts}
\usepackage{xcolor}

\title{Starting inflation from inhomogeneous initial conditions with momentum
}

\author[a,b]{Maxence Corman}
\author[a]{William E. East}

\affiliation[a]{
   Perimeter Institute for Theoretical Physics,
   Waterloo, Ontario N2L 2Y5, Canada
}
\affiliation[b]{
   Department of Physics and Astronomy,
   University of Waterloo,
   Waterloo,
   Ontario N2L 3G1,
   Canada
}

\emailAdd{mcorman@perimeterinstitute.ca}
\emailAdd{weast@perimeterinstitute.ca}

\abstract{ 
   We investigate the circumstances under which cosmic inflation can arise from very
   inhomogeneous initial conditions using numerical relativity simulations.
   Previous studies have not considered cases with non-zero momentum density
   due to technical challenges with solving the coupled Einstein constraint
   equations. Here we address these, introducing and comparing several
   different ways of constructing cosmological initial conditions with
   inhomogeneous scalar field and time derivative profiles.  We evolve such
   initial conditions with large inhomogeneities in both  single- and two-field
   inflationary models. We study cases where the initial gradient and
   kinetic energy are much larger than the inflationary energy scale, and black
   holes can form, as well as cases where the initial scalar potential energy
   is comparable, as in scenarios where inflation occurs at nearly Planckian
   densities, finding large-field inflation to be generally robust.  We
   consider examples of initial conditions where a large scalar field velocity
   towards non-inflationary values would prevent inflation from occurring in
   the homogeneous case, finding that the addition of large gradients in the
   scalar field can actually dilute this effect, with the increased expansion
   and non-vanishing restoring force leading to inflation.
}

\begin{document}
\maketitle
\flushbottom

\section{Introduction}
\label{sec:intro}
Cosmic inflation
\cite{Starobinsky:1980te,Guth:1980zm,Linde:1981mu,Albrecht:1982wi,
Linde:1983gd} has been proposed as a solution to some of the fine-tuning
problems of Standard Big Bang cosmology, namely, the horizon and flatness
problems.  But in order for inflation to be a successful explanation for the
observed homogeneity of the universe on large scales, it should be able to
arise from generic, inhomogeneous initial conditions.  Inhomogeneities will not
necessarily spoil inflation in models where inflation naturally begins at
nearly Planckian densities,
\cite{Linde:1983gd,Linde:1984ir,Linde:1985ub,Linde:2014nna, Linde:2017pwt},
though this has not be studied in detail.  The most recent observations of the
cosmological microwave background motivate studying of the problem of initial
conditions in low energy scale inflation, in which the potential energy density
is much below the Planckian scale
\cite{Planck:2018jri,BICEP:2021xfz,Linde:2014nna,Kallosh:2014xwa,Linde:2017pwt,Kallosh:2021mnu}.
Here, we focus on the effects of large inhomogeneities on the onset of
inflation, both in scenarios where it occurs at nearly Planckian and sub-Planckian energy scales, using
evolutions in fully nonlinear general relativity. 

This question has been studied using tools from numerical relativity in a number of papers 
\cite{Kurki-Suonio:1987mrt,
Brandenberger:1988ir,Goldwirth:1989vz,Goldwirth:1989pr,Goldwirth:1990pm,
Laguna:1991zs,Goldwirth:1991rj,Kurki-Suonio:1993lzy,Deruelle:1994pa,East:2015ggf,
Clough:2016ymm,Braden:2016tjn,Clough:2017efm,Aurrekoetxea:2019fhr,Joana:2020rxm},
complementing work evolving inhomogeneous fields on homogeneous spacetimes~\cite{Albrecht:1985yf,Albrecht:1986pi,Kung:1989xz,Alho:2011zz,Easther:2014zga}, 
and using analytic techniques~\cite{1985MNRAS.216..395B,Kleban:2016sqm,Creminelli:2020zvc}
(see \cite{Brandenberger:2016uzh} for a short review on the topic). 
Focusing on more recent work, \cite{East:2015ggf} showed that large field inflation is robust 
to simple inhomogeneous initial conditions even when the initial gradient energy 
is many orders of magnitude larger than the vacuum energy density,
provided that the universe is initially expanding
everywhere and that the scalar field range remains within the slow-roll regime. 
Reference \cite{East:2015ggf} also included cases with large inhomogeneities 
that give rise to the formation of black holes, but showed that inflation 
can succeed even then. 
This happens because while the overdense scalar field regions collapse into black holes, 
the underdense regions evolve into voids that become dominated by the inflationary 
potential energy at later times such that inflation may begin. 
This line of research was then extended in two ways: 
\cite{Clough:2016ymm,Aurrekoetxea:2019fhr,Joana:2020rxm,Joana:2022uwc} 
expanded the classes of inflationary models under investigation and 
\cite{Clough:2016ymm,Clough:2017efm,Joana:2020rxm} the classes of inhomogeneities. 
In \cite{Clough:2016ymm}, inhomogeneities in both the scalar field profile 
and the extrinsic curvature were explored. In particular, the initial expansion was assumed to take 
the following simple form $K(\vec{x}) = - C \phi(\vec{x}) + K_0$ where $C>0$ is a free parameter 
and $K_0$ is an integration constant and the initial velocity of scalar field was given 
by some constant such that the momentum constraint is trivially satisfied. 
In this ansatz, 
the initial hypersurface contains regions of local expansion and contraction. 
It was shown that as long as the spacetime is on average initially expanding, 
then inflation will occur in some patch even if other parts of the spacetime might collapse. 
Reference~\cite{Clough:2017efm} investigated inhomogeneities in the transverse traceless 
part of the extrinsic curvature, $A^{TT}_{ij} \neq 0$, and found that these initial perturbations,
roughly corresponding to gravitational radiation, 
initially reduce the total number of $e$-folds as the amplitude is increased, but that this reduction 
saturates and that in general the amplitude of the scalar perturbations remains the main driver 
of the onset of inflation. The effect of initial inhomogeneities in the scalar kinetic energy, but with
constant field profile, such that the scalar momentum density was initially zero, was studied in~\cite{Joana:2020rxm}.

In order to carry out such evolutions of various classes of initial conditions
to determine whether they eventually lead to an exponentially expanding
spacetime, one must begin with initial data that satisfies the constraint part
of the Einstein equations.  In practice, this requires specifying the values of
various metric and matter components on the initial time slice in such a way
that they satisfy the Hamiltonian and momentum constraint equations.  However,
all of the abovementioned studies are similar in that they rely on special
choices of initial conditions that ensure the momentum constraint is trivially
satisfied. The addition of momentum in the initial conditions is not only
natural, but also interesting to pursue since, if the initial velocity of the
inflaton were large enough, it could prevent the onset of inflation.  Solving
the momentum constraint, in addition to the Hamiltonian constraint is highly
non-trivial, not only because it requires solving three additional coupled
non-linear elliptic equations, but also because it is challenging to separate
freely specifiable versus constrained degrees of freedom in a manner where the
underlying physical interpretation of the free data is transparent, while also
ensuring that a unique solution exists. If we over-restrict the system, no
solutions to the constraint equations will exist, also known as the problem of
\emph{existence}. On the other hand, if we do not restrict the system
sufficiently, there may be multiple nearby solutions, in which case our
numerical solver might not converge to a single one. This so-called problem of
local \emph{uniqueness} is very relevant in cosmology \cite{Garfinkle:2020iup},
and has implications for how matter source terms in the momentum and
Hamiltonian constraint are specified, as discussed below.  In addition, a
standard trick to choose which variables to set, and which to solve for, while
ensuring the solution is unique, is to perform a conformal decomposition of the
energy and momentum density. However, this does not extend straightforwardly to
scalar fields since the energy and momentum density are functions of the scalar
field and its time and spatial derivatives.

In this work, we solve both the Hamiltonian and momentum constraint equations
for non-trivial inhomogeneities in the momentum density of the scalar field,
considering scenarios where the time derivative of scalar field has a large homogeneous value, 
as well as
scenarios where it has large spatial variations.
We solve the equations
using the conformal thin-sandwich (CTS) formalism \cite{York:1998hy}, 
making using of the code described in \cite{East:2012zn}, but without using the usual conformal rescaling
of the matter terms. 
We find that our new scheme gives greater control over the initial conditions
in the matter sector.  We then study the evolution of several classes of
initial conditions that have non-trivial inhomogeneities in the scalar field
momentum density and have not been previously studied in the literature.  We
consider scenarios where the length scale of the inhomogeneities is comparable
to the initial effective Hubble radius and  (i) the initial scalar gradient and
kinetic energy are comparable and much larger than the inflationary energy
scale; (ii) the initial scalar gradient, kinetic energy, and potential energy
are all comparable (as might arise in large field inflationary models where
inflation begins at nearly Planckian densities); and (iii) in two-field
inflationary models where the scalar fields are non-interacting.  Our results
for (i) are, broadly speaking, in agreement with previous analyses that assumed
a vanishing initial velocity profile for scalar field.  However, we also show
that when the initial kinetic energy of scalar field is such that a homogeneous
universe will fail to inflate, the addition of gradient energy can slow the
scalar field before it reaches the end of the inflationary plateau.  We
attribute this to an increased initial expansion rate and non-negligible
pullback force due to the presence of gradients in the inflaton.  Our results
suggest that large gradients can mitigate the disruptive effect of a non-zero
initial scalar velocity profile.  Our results for (ii) show that in cases where
the kinetic, gradient, and potential energy are all comparable, the universe
will rapidly transition to exponential expansion, typically without forming
collapsing regions.  Finally, we extend our methods to the study of the effects
of adding inhomogeneities to cosmological scenarios where the universe undergoes two
stages of inflation. 
Taken together, our
results suggest inflation can arise from highly inhomogeneous conditions.

The remainder of this paper is as follows. We discuss the inflationary models we use in 
section~\ref{sec:theory}. Our approach to solving the constraint equations on the initial
time slice is outlined in section~\ref{sec:metric} and ~\ref{sec:matter}. 
We comment on the existence and uniqueness of our solutions in ~\ref{sec:max_princ}.
Our numerical
methods and diagnostics for evolving the Einstein equations and matter are described in 
section~\ref{sec:numerical_method}. Our results are presented in 
section~\ref{sec:results}, and we conclude in section~\ref{sec:conclusion}. 
In appendix~\ref{app:numerics}, we discuss our numerical implementation in more detail,
while in 
appendix~\ref{app:MCTS}, we provide more details on the construction of our initial data.

We work in four spacetime dimensions, with metric signature
$(-+++)$; we use lower-case Greek letters ($\mu,\nu,...$)
to denote spacetime indices and Latin letters ($i,j,k,...$, although $t$
is reserved for the time coordinate index)
to denote spatial indices.
The Riemann tensor is
$R^{\alpha}{}_{\beta\gamma\delta}
=
\partial_{\gamma}\Gamma^{\alpha}_{\beta\delta}-\cdots
$.
We use units with $M_{P}\equiv 8 \pi G = c = 1$.

\section{Theory}\label{sec:theory}
In this paper, we consider inflationary theories with one or two canonical 
scalar fields, $\phi$
and $\theta$, both minimally coupled to Einstein gravity, such that the action
is given by 
\begin{equation}\label{eq:action}
S=\int d^4x \ \sqrt{-g} \left(\frac{1}{2} R - 
	g^{\mu\nu} \partial_{\mu} \phi \partial_{\nu} \phi 
	- g^{\mu\nu} \partial_{\mu} \theta \partial_{\nu} \theta -2 V (\phi,\theta) \right)
\end{equation}
where $g$ is the determinant of the spacetime metric, 
$R$ the Ricci scalar and $V(\phi,\theta)$ the potential energy function. 
In most of the cases, we will consider single-field inflation by setting $\theta=0$.
In all cases, we assume that the scalar fields do not interact, such that 
$V(\phi,\theta) = V_{\phi}(\phi)+V_{\theta}(\theta)$.
The Einstein equations that follow from the action \eqref{eq:action}, are then
\begin{equation}\label{eq:einstein}
    G_{\mu\nu}=R_{\mu\nu}-\frac{1}{2} R g_{\mu\nu}=T_{\mu\nu} 
\end{equation}
where the stress-energy tensor is
\begin{equation}
	T_{\mu\nu}= T_{\mu\nu}^{(\phi)} + T_{\mu\nu}^{(\theta)}
\end{equation}
with
\begin{equation}\label{eq:T}
	T_{\mu\nu}^{(\phi)}=g_{\mu\nu} \left[- \partial_{\alpha} \phi \partial^{\alpha} \phi 
	-2 V_{\phi}(\phi)\right] +2 \partial_{\mu} \phi \partial_{\nu} \phi \ , 
\end{equation}
and similarly for $\theta$.
The equation of motion for each individual field is then
\begin{equation}\label{eq:eom_phi}
	\Box \phi = V_{\phi}', \ \ \ \	
	\Box \theta = V_{\theta}'
\end{equation}
where $'$ is used to denote
the partial derivative with respect to corresponding scalar field. 
In general, each field will undergo inflation when its potential energy density is 
greater 
than the sum of its kinetic and gradient energy density. Assuming the universe is spatially 
homogeneous, $\phi$ will undergo inflation if $V_{\phi}>0$ and the slow-roll parameters satisfy:
\begin{equation}
	\epsilon_{\scaleto{V}{3.5pt}} =\left(\frac{V_{\phi}'}{2 V_{\phi}}\right)^2 
	\ll 1, \ \ \ \ \eta_{\scaleto{V}{3.5pt}} =\frac{1}{2} \left|\frac{ 
	V_{\phi}''}{V_{\phi}}\right|  \ll 1 ,
\end{equation}
in which case the field is slowly rolling. 
The condition $\epsilon_{\scaleto{V}{3.5pt}}  \ll 1$ implies that the potential energy 
density driving inflation is roughly constant, 
$\rho_{V_{\phi}} \equiv 2 V_{\phi} \approx \mathrm{const.}$, 
while $\eta_{\scaleto{V}{3.5pt}} \ll 1$ ensures that inflation persists for a number of $e$-folds.

\subsection{Initial conditions: metric}\label{sec:metric}
We wish to specify initial data on a spacelike hypersurface $\Sigma_t$ parameterized by $t$
that is consistent with the Einstein equations.
We formulate the problem using the Arnowitt-Deser-Misner 
(ADM) formalism, and decompose the metric 
into a spatial metric $\gamma_{ij}$, lapse $N$, and shift vector $\beta^i$ as 
\begin{equation}
ds^2= -N^2 dt^2 + \gamma_{ij} (dx^i +\beta^i dt)(dx^j +\beta^j dt) \ ,
\end{equation}
and write the extrinsic curvature as 
\begin{equation}
	K_{ij} = - \frac{1}{2}\mathcal{L}_{\vec{n}} \gamma_{ij} \ ,
\end{equation}
where the Lie derivative $\mathcal{L}$ is taken with respect to the timelike unit normal to slices 
of constant coordinate time $n^{\mu}=\left(1/N,-\beta^i/N\right)$. 
The initial data for the metric and matter sector cannot be freely and independently
specified, but must satisfy the momentum and Hamiltonian constraint
equations 
\begin{eqnarray}
	^{(3)} R +K^2 -K^{ij}K_{ij}=2 \ \rho \label{eq:hamiltonian} \ , \\
	D_j K^{ij} -D^i K =  \ p^i \label{eq:momentum} \ ,
\end{eqnarray}
where $K=\gamma^{ij}K_{ij}$, $^{(3)}R$, and $D_i$ are, respectively, the extrinsic curvature scalar, Ricci scalar, 
and covariant derivative associated with $\gamma_{ij}$, 
$\rho \equiv n^{\mu}n^{\nu} T_{\mu\nu}$ is the energy density, and 
$p^i \equiv -\gamma^{i\mu}n^{\nu} T_{\mu\nu}$ is the momentum density as measured 
by an Eulerian observer with four-velocity $n^{\mu}$. 

In the 3+1 decomposition, initial data for the Einstein and matter equations are
then a set of 20 functions representing $N$, $\beta^i$, $\gamma_{ij}$, $K_{ij}$, $\rho$, and $p^i$
on the initial time slice that together satisfy the constraint equations.

We use a modified version of the CTS method \cite{York:1998hy}, implemented in 
\cite{East:2012zn}, as a prescription to separate freely specifiable from 
constrained degrees of freedom in such a way that the underlying physical interpretation
of the free data is transparent. 
We outline the key features below, and refer
the reader to \cite{East:2012zn} for more details.
In the CTS formalism, the approach is to perform a conformal decomposition of 
the spatial metric and
extrinsic curvature in order to introduce quantities that can be specified in a more
well-motivated way. Introducing the conformal factor $\Psi$, we can define
a number of conformal quantities as
\begin{eqnarray}\label{eq:conformal_eqns}
\tilde{\gamma}_{ij} &\equiv & \Psi^{-4} \gamma_{ij}  \\
\hat{A}^{ij} &\equiv & \Psi^{10} \left( K^{ij} -\frac{1}{3} K \gamma^{ij}\right) \nonumber \\
&=&\frac{1}{2\tilde{N}} 
	\left[\tilde{\dot{\gamma}}^{ij} + (\mathbb{L} \beta)^{ij}\right]
\end{eqnarray}
where the time-derivative of the conformal metric
$\dot{\tilde{\gamma}}^{ij} \equiv \Psi^4 \left( \dot{\gamma}^{ij} 
-  \frac{1}{3}\gamma^{ij}\gamma_{kl}\dot{\gamma}^{kl} \right)$ 
is traceless by construction. The conformal lapse $ \tilde{N}$ is related to the
lapse by  $\tilde{N} \equiv \Psi^{-6}N$, while
$\tilde{R}$ and $\tilde{D}$ are the Ricci scalar and covariant derivative 
with respect to $\tilde{\gamma}_{ij}$, and $\mathbb{L} $ is the corresponding conformal
Killing operator, defined by
\begin{equation}
(\mathbb{L} \beta)^{ij} \equiv \tilde{D}^i \beta^j +\tilde{D}^j \beta^i -\frac{2}{3} \tilde{\gamma}^{ij} \tilde{D}_k \beta^k 
\ .
\end{equation}
With these definitions, the constraint equations become an elliptic condition for $\Psi$,
and three elliptic equations for $\beta^i$:
\begin{eqnarray}
	\tilde{D}_i\tilde{D}^i\Psi-\frac{1}{8}\tilde{R}\Psi + 
	\frac{1}{8}\hat{A}_{ij}\hat{A}^{ij}\Psi^{-7}-\frac{1}{12}K^2\Psi^5 
	+\frac{1}4{} \ \rho \Psi^{5} &=& 0, \label{eq:cts_hamiltonian} \\
	\tilde{D}_j\hat{A}^{ij} - \frac{2}{3}\Psi^6\tilde{D}^iK 
	-  \ \Psi^{10} p^i &=& 0 \label{eq:cts_momentum} \ .
\end{eqnarray}
In the CTS formalism, one then usually proceeds by specifying initial data for the free data
\begin{equation}
\tilde{\gamma}_{ij}, \ \ \ \dot{\tilde{\gamma}}_{ij}, \ \ \ K, \ \ \ \tilde{N}, 
	\ \ \ \rho, \ \ \ p^i
\end{equation}
(or some conformally rescaled version of $\rho$ and $p^i$), 
and solving the four elliptic equations for the shift $\beta^i$ and conformal factor $\Psi$.
We emphasize that in the CTS method, one is in principle free to choose
any values for the free data for which a solution can be found. We now outline and motivate
our choice of free data
for the case of inhomogeneous conditions in an initially expanding universe filled with
two (unless otherwise stated) scalar fields and periodic boundary conditions
\footnote{
Our simulation domain is periodic with coordinate length $L$, the largest
wavelength of the perturbations in the scalar field, in each direction.  This
can either be seen as a convenient boundary condition for an infinite universe,
or in scenarios where inflation starts slightly below Planckian densities, as
describing a universe with toroidal topology. 
}.

For simplicity, we choose
the spatial metric on the initial time slice to be conformally 
flat, $\gamma_{ij} \equiv \Psi^4 \tilde{\gamma}_{ij} = \Psi^4 \delta_{ij}$, $\tilde{N}=1$, and
set the trace of the extrinsic curvature $K$ 
to be constant, while the transverse-traceless part $\dot{\tilde{\gamma}}^{ij}$ is set to zero 
(loosely this is equivalent
to setting the gravitational wave background to zero). 
In principle, $K$ is a free parameter, representing a uniform expansion rate across the 
initial hypersurface. 
However, we also have to satisfy
an integrability condition for the Hamiltonian constraint
\begin{equation}\label{eq:int_H}
	\mathcal{I}_{\mathcal{H}} \equiv \int \left( \tilde{D}_i\tilde{D}^i\Psi-\frac{1}{8}\tilde{R}\Psi + 
	\frac{1}{8}\hat{A}_{ij}\hat{A}^{ij}\Psi^{-7}-\frac{1}{12}K^2\Psi^5 
	+\frac{1}4{} \ \rho \Psi^{5}   \right) dV = 0 \ .
\end{equation}
For a periodic domain, if we approximate the conformal factor to be roughly unity, this requires
$K^2/3$ to be close to the 
initial energy density averaged over the initial hypersurface. For two scalar fields, this 
reduces to
\begin{equation}\label{eq:trK}
	K=-\left[\frac{3}{\rm Vol} \int \left( \left(\partial_t{\phi}\right)^2+
	\tilde{\gamma}^{ij}\partial_i \phi \partial_j \phi +2 V_{\phi}(\phi)
	+\left(\partial_t{\theta}\right)^2+\tilde{\gamma}^{ij}\partial_i \theta \partial_j \theta 
	+2 V_{\theta}(\theta)
    \right) dV \right]^{1/2} 
\end{equation}
where $\phi$ and $\theta$ are the values of scalar fields on the initial time slice and
$\partial_t \phi$ and $\partial_t \theta$ are specified up to some conformal factor.
The minus sign gives us an initially uniformly expanding universe with Hubble parameter $H_0=-K/3$ 
(a positive $K$ would imply a contracting universe). In the special case of a homogeneous 
scalar field profile, this choice of initial data gives a Friedman-Lemaitre-Robertson-Walker
(FLRW) solution.

We now describe the choice of free data in the matter sector.
We will restrict the discussion to a single field, say $\phi$, but keep in mind
that the same definitions and assumptions apply to the other field $\theta$.
\subsection{Initial conditions: matter}\label{sec:matter}
In the case of a scalar field, 
the quantities to be specified on the three
dimensional spatial hypersurface are the scalar field $\phi$ and its 
time derivative $\partial_t \phi$.
We thus rewrite the energy density at some time and any point 
in the hypersurface as the sum of the kinetic, gradient, and potential energy densities,
\begin{equation}\label{eq:energy_density}
	\rho = \eta^2 +\gamma^{ij} \partial_i \phi \partial_j \phi +2 V_{\phi}(\phi)
\end{equation}
and the momentum density as
\begin{equation}\label{eq:momentum_density}
p^i = -2 \ \eta \ \gamma^{ij} \partial_j \phi
\end{equation}
where $\eta$ is the conjugate momentum
\footnote{The conjugate momentum can be interpreted as the velocity of scalar field as seen 
by a fiducial observer whose worldline is orthogonal to the constant time hypersurfaces 
$\Sigma_t$, i.e., an observer with four-velocity $n^{\mu}$. }
defined by
\begin{equation}\label{eq:stkin}
	\eta \equiv n^\mu \nabla_{\mu} \phi = \frac{1}{N} \left(\partial_t \phi -\beta^i \partial_i \phi \right).
\end{equation}
Clearly, the technical challenge here is to choose which quantities to specify in such a way that
not only does a unique solution exist, but also such that the values for the quantities can be physically motivated.
The fundamental quantities we freely specify are the initial scalar field profile
and a non-trivial velocity profile defined below.
We comment on the existence and uniqueness of our solutions in section \ref{sec:max_princ}.
The inhomogeneous initial conditions for the scalar field profile are chosen to be 
such that on the initial time slice,
\begin{equation}\label{eq:phi0}
\phi(t=0,\vec{x})=\phi_0 +  \delta \phi_x \sin (k x) + 
	\delta \phi_y \sin (k y)+ \delta \phi_z \sin (k z) 
\end{equation}
where $\vec{x}=(x,y,z)$ is the spatial coordinate of hypersurface labelled by the time 
coordinate $t$, $k=2 \pi/L$ is the wavenumber and 
$\delta \vec{\phi}=(\delta \phi_x, \delta \phi_y, \delta \phi_z)$ is a measure of the amplitude of 
the initial inhomogeneities which in general can be different in each direction, but
we will only consider $ \delta \phi =\delta \phi_x= \delta \phi_y= \delta \phi_z$ . 
The maximum total amplitude of the fluctuations about $\phi_0$ is then 
$\Delta \phi = \sqrt{\delta \phi_x^2+\delta \phi_y^2+\delta \phi_z^2}= \sqrt{3} \delta \phi$. 
For simplicity, we do not consider different amplitudes in the different
spatial directions, or additional scalar field variations at smaller wavelengths. In
\cite{East:2015ggf}, including these was found not to qualitatively affect the results.
We fix the (coordinate) length of the simulation domain in each direction to be
equal to the wavelength $L$, and consider various ratios of this lengthscale to
the initial Hubble scale (including gradient and all other energy contributions). 
The initial conditions of the scalar field depend on the amplitude of the inhomogeneities 
$\delta \phi$ and the potential $V(\phi_0)$ which sets the background inflationary Hubble scale. 

We explore the effects of inhomogeneities on
both single-field and two-field inflationary models.
When studying single-field inflation, we 
consider two types of inflationary potentials, both of which 
are large field models, i.e.,
the inflaton needs to traverse a super-Planckian range in field space for there
to be a sufficient period of inflation.
The first potential we consider is a simple quadratic potential,
\begin{equation}\label{eq:Vm}
	V(\phi)=m^2 \phi^2 \ .
\end{equation}
Although strongly disfavored by the most recent observations \cite{Planck:2018jri,BICEP:2021xfz},  
we do not expect our conclusions to be dependent on the shape 
of the potential, since the key feature is the flatness of potential
and super-Planckian distance in field space to the minimum of potential
(see \cite{Aurrekoetxea:2019fhr} for a study of the effects of 
the potential shape on inhomogeneous inflation). 
To confirm this, we do, however, also look at the T-shaped potential,
\begin{equation}\label{eq:NOTCH}
V(\phi) = 3m^2 \alpha \tanh^2{\left(\frac{\phi}{\sqrt{6 \alpha}}\right) }
\end{equation}
describing a class of so-called $\alpha$-attractors, motivated by supergravity and string theory 
\cite{Kallosh:2013hoa,Kallosh:2013yoa} and consistent with the most recent BICEP/KECK data 
\cite{Kallosh:2021mnu}.
The two-field model we will consider is given by
\begin{equation}\label{eq:2field}
	V(\phi,\theta) = M^2 \phi^2 + 3 m^2 \alpha \tanh^2{\left(\frac{\theta}{\sqrt{6 \alpha}}\right) }
\end{equation}
where $M \gg m$. Here, inflation can start at $M^2 \phi^2 = O(1)$ if all gradient and kinetic 
terms are much smaller than $V_{\phi}(\phi)$~\cite{Linde:2017pwt}. 
When the period of inflation driven by $\phi$ ends, 
a second stage of inflation driven by $\theta$ begins.
Our choice of two-field model is motivated by the recently proposed
$\alpha$-attractor generalization \cite{Kallosh:2022ggf,Braglia:2022phb}
of the hybrid inflation scenario \cite{Linde:1993cn}.
Although this scenario describes an inflaton field $\theta$, interacting with a Higgs-type
field $\phi$, one can show that
the fields become decoupled in the large field limit, 
in which case the inflationary model reduces to the two-field model studied here.

Going back to the initial conditions on the velocity profile the scalar field,
we find three distinct ways of specifying well motivated initial data, which we now
describe.
\begin{enumerate}
\item \textit{Uniform initial time derivative of scalar field}.\\
The first method is to specify the time derivative of the scalar field, which for simplicity we
assume to be constant on the numerical domain
\begin{equation}\label{eq:PhistDotCnst}
\partial_t \phi (t=0,\vec{x})= \omega  \delta\phi
\end{equation}
where $\delta \phi \sim O(1)$. 
The sign of the parameter $\omega$ determines whether the scalar field moves up or down the inflationary
potential. In particular, when $\omega < 0$, the field is driven towards the minimum of the
inflationary potential, which could cause an early 
end to inflation. We will thus give particular attention to this scenario. 

We emphasize that, although the time derivative of the scalar field is constant, after the Hamiltonian
and momentum constraints are solved for $\Psi$ and $\beta^i$, the reconstructed physical energy 
and momentum density are inhomogeneous. While the reconstruction of the physical energy 
densities is unambiguous, the lack of ability to specify a particular configuration for
$\rho$ and its individual components is inconvenient. This motivates another way of 
specifying initial data.
\item \textit{Spatially varying conjugate momentum: conformal rescaling}.\\
Here, the trick is to specify a rescaled conjugate momentum, defined as
\begin{equation}\label{eq:RF_eta}
	\tilde{\eta} = \Psi^2 \eta.
\end{equation}
This ensures that if the conformal gradient energy,
$\tilde{\gamma}^{ij} \partial_i \phi \partial_j \phi$ and kinetic energy $\tilde{\eta}^2$
have a chosen ratio, then so will the physically reconstructed quantities following the solution
of the constraints. In other words, the ratio of the initial kinetic and gradient energies 
is not modified by solving for the shift and conformal factor in CTS formalism. 
The ansatz for $\tilde{\eta}$ is chosen to be spatially varying, although with fixed
negative sign on the entire domain
\begin{equation}\label{eq:RF_stkin}
\tilde{\eta}=- k  \delta \phi \sqrt{  \cos^2(k x)+ \cos^2(k y)+\cos^2(k z) }.
\end{equation}
such that the scalar field initially moves down the inflationary potential.
We note that this initial data breaks the symmetry of the energy density about the origin.
\item \textit{Spatially varying conjugate momentum: analytical solution}.\\
Finally, instead of solving the momentum and Hamiltonian constraints using the 
CTS formalism, a third approach is to, following~\cite{Garfinkle:2008ei,Xue:2013bva},
choose initial data such that the momentum constraint
is analytically satisfied. We then use the Hamiltonian constraint to solve for the conformal factor. 
We outline the main assumptions of this construction here, and 
refer the reader to appendix~\ref{app:MCTS} for more details. 

The initial data for the spatial metric and extrinsic curvature scalar
is specified the same way as above, but the lapse and shift are chosen to be 
$N=1$ and $\beta^i=0$. The initial data for the scalar field profile is obtained by
specifying $\phi(t=0,\vec{x})$ according to \eqref{eq:mom_triv} below, and the conjugate momentum 
by specifying
\begin{equation}\label{eq:eta_trivial}
\hat{\eta} = \Psi^6 \eta= \Psi^6 \partial_t \phi 
\end{equation}
where it is important to note that $\hat{\eta}$ has a different scaling
with the conformal factor from $\tilde{\eta}$ introduced 
above \eqref{eq:eta_trivial}
and that the last step assumes $N=1$ and $\beta^i=0$.
Given those assumptions, the momentum constraint is then solved by the following ansatz
\begin{eqnarray}\label{eq:mom_triv}
\hat{\eta}(x)&=&\frac{1}{\sqrt{2}}\left[\hat{\eta}_0 +\delta \hat{\eta}_x \cos (kx)
	+\delta \hat{\eta}_y \cos (ky) +\delta \hat{\eta}_z \cos (kz)\right], \\
\phi(x) &=& \frac{1}{\sqrt{2}}\left[\hat{\phi}_0 +\delta \hat{\phi}_x \cos (kx)+\delta \hat{\phi}_y
	\cos (ky) +\delta \hat{\phi}_z \cos (kz)\right]
\end{eqnarray}
and a particular solution for $\hat{A}^{ij}$ given in eqn.~\eqref{eq:Aij_mcts}.
This allows the reconstructed $\partial_t{\phi}(t=0,\vec{x})$ 
to take both positive and negative values in the domain.
For this study, we will further assume $\delta \hat{\eta} = \delta \hat{\eta}_x =\delta \hat{\eta}_y=
\delta \hat{\eta}_z$ and
$\delta \hat{\phi} = \delta \hat{\phi}_x = \delta \hat{\phi}_y = \delta \hat{\phi}_z$.
\end{enumerate}
\subsection{Local uniqueness and existence}\label{sec:max_princ}
We now briefly discuss how the different ways of specifying scalar field
initial data relate to considerations of how the Hamiltonian and momentum
constraint couple, and issues of local uniqueness.\footnote{We do not go into
any detail here. See,
e.g,~\cite{Murchadha:1974,Smarr:1979ofa,Walsh:2006au,Bartnik:2002cw} for
reviews on the uniqueness of constraint equations.} First, we recall the more
familiar case with fluid matter initial data, in order to compare and
contrast to the scalar field case.  In the former, where the stress tensor
is an algebraic function of the fluid variables, one typically specifies a
conformal energy density $\tilde{\rho} \equiv \Psi^n \rho$ and a conformal
momentum density $\tilde{p}^i \equiv \Psi^{10}{p}^i$. For the energy density, a
value $n>5$ is chosen based on uniqueness considerations, which we discuss
below, while the exponent of the conformal momentum density is chosen to remove
the $\Psi$ dependence in the momentum equation~\eqref{eq:cts_momentum}
\cite{York:1998hy}. In the special case where $K$ is chosen to be constant, the momentum
constraint then becomes independent of $\Psi$, and the constraint equations
decouple in the sense that one can first solve the momentum constraint for
$\beta^i$, and then using the solution for $\beta^i$, solve the Hamiltonian
constraint for the conformal factor. 

In contrast, in the scalar field case matters are complicated by the fact that
the energy and momentum depend on the gradient of the scalar field. As can be
seen from eq.~\eqref{eq:energy_density}, specifying a conformal energy density would involve
solving an additional nonlinear partial differential equation to determine the scalar field, and we are
typically interested in specifying the initial scalar field profile as the
quantity of physical interest anyway.  However, noting that $\tilde{p}^i=-2\Psi^6 \eta
\tilde{\gamma}^{ij} \partial_j \phi$ is fixed by choosing $\phi$ (and hence $\partial_j
\phi$) and say $\hat{\eta}\equiv \eta \Psi^6$ at $t=0$, then this choice of scalar field
initial data achieves the same outcome as above, and (since we always choose
$K$ to be constant here) the momentum constraint decouples from the Hamiltonian
constraint. This is the case considered above when constructing initial data such that the momentum constraints 
are analytically satisfied \eqref{eq:eta_trivial}, and is also the case 
where there are uniqueness results in the mathematical literature for a scalar field on a compact manifold~\cite{Choquet-Bruhat:2006khn}.
As discussed in Ref.~\cite{Garfinkle:2020iup}, 
issues with the non-uniqueness of the momentum constraint can arise in this case when the kernel of the operator $\partial_j \mathbb{L}$ is non-trivial,
which occurs with typical choices of initial data (e.g. a conformally flat metric). 
These are side-stepped with the analytical solution used here.
Fixing $\partial_t \phi$, as in eq.~\eqref{eq:eta_trivial}, also removes the dependence
of the momentum constraint on $\Psi$, since 
$\Psi^6\eta = \left(\partial_t \phi +\beta^i\partial_i \phi\right)/\tilde{N}$.
However, it introduces an extra term involving $\beta^i$, so that the previous analyses do not directly apply.
For the conformal rescaling~\eqref{eq:RF_eta} that fixes the ratio of kinetic to gradient energy, the momentum constraint
does not decouple and, again, previous analyses do not apply.

Finally, we consider the local uniqueness of the Hamiltonian constraint.  For
simplicity, we will ignore the momentum constraint and restrict to the choices
for the metric free data discussed in Sec.~\ref{sec:metric} (e.g. conformally
flat, and so on). Say we have some solution $\Psi_0$ to the CTS form of the
Hamiltonian constraint~\eqref{eq:cts_hamiltonian}, and we are interested in
nearby solutions $\Psi = \Psi_0 + u$ where $u$ is small compared to $\Psi_0$.
Linearizing, we have 
\begin{equation}
    \left[\tilde{D}_i \tilde{D}^i -q(\vec{x})\right] u =  0
\end{equation}
where, in terms of the functional derivative of the kinetic energy term, 
\begin{equation}\label{eq:specific_q}
	q(\vec{x}) = \frac{7}{8} \Psi_0^{-8} 
	(\mathbb{L}\beta)_{ij}(\mathbb{L}\beta)^{ij}
	+\frac{5}{12} \Psi_0^4 K^2
    -\frac{5}{4}\Psi_0^4 (2 V_{\phi}) -\frac{1}{4}\tilde{\gamma}^{ij}
	\partial_i \phi \partial_j \phi
    -\frac{1}{4} \left. \frac{\delta (\eta^2 \Psi^5)}{\delta \Psi} \right\rvert_{\Psi = \Psi_0} \ .
\end{equation}
If $q(\vec{x})>0$, then, using the maximum principle, one can show that $u=0$
everywhere, and $\Psi_0$ is (locally) unique~\cite{choquet1992solutions,Choquet-Bruhat:1999gli}.  For the case of fluid matter
mentioned above, the choice of $\tilde{\rho}$ is motivated by ensuring that
$\tilde{\rho}$ makes a positive contribution to $q$.  In the scalar case,
clearly, this will not be true in general.  
When fixing $\hat{\eta}$ or $\partial_t \phi$,
the functional derivative term in eq.~\eqref{eq:specific_q} becomes
$7\Psi_0^{-8}\hat{\eta}^2/4$,
and hence makes a positive contribution to $q$. When
specifying $\tilde{\eta}$, the contribution is $-\tilde{\eta}^2/4$, and hence negative (as noted
above, in this case the momentum constraint does not decouple as well). 
In either case, for the parameters considered here, we find that $q(\vec{x})>0$.
In part, this can be attributed to our choice of $K$ through eq.~\eqref{eq:trK}. 
A sufficient condition for $q>0$ is that 
\begin{equation}
	\langle  2V_\phi+\tilde{\gamma}^{ij}\partial_i \phi \partial_j \phi +\hat{\eta}^2 \rangle \gtrsim  
	\left(2V_\phi+\frac{1}{5}{\gamma}^{ij} \partial_i \phi \partial_j \phi - \frac{7}{5}{\eta}^2\right)
\label{eq:posq_cond}   
\end{equation}
when fixing $\hat{\eta}$ or $\partial_t \phi$, where here $\langle \ldots \rangle$ denotes the coordinate volume average. 
When fixing $\tilde{\eta}$ instead, the last term becomes $+{\eta}^2/5$.
In all the cases we consider in this study, we find $\Psi_0 \sim 1$, and the gradient and kinetic energy terms, which appear with smaller
prefactor on the right-hand side of~\eqref{eq:posq_cond} compared to the left-hand side,
are approximately equal in magnitude, and either larger, or approximately equal, to the scalar potential energy term.
This argument can also be straight-forwardly extend to multiple non-interacting scalar fields, in which
case the kinetic, gradient and potential energy densities in \eqref{eq:posq_cond} would
simply be the sum of energy densities of individual fields.

Though the above discussion provides some motivation and guidance, in the end
there is no mathematical result guaranteeing the existence and uniqueness of
solutions to the constraint equations for all the cases we consider.  
However,
the important thing for the purposes of this study is that we find we are able
to numerically construct convergent solutions. In fact, there are several
examples in the literature where the constraint equations have been solved
without issue when it was known that more than one solution
exists~\cite{Yoshida:1997qf,Gourgoulhon:2000nn,Yo:2004ng,Pfeiffer:2005jf,Baumgarte:2006ug}.

\subsection{Numerical implementation}\label{sec:numerical_method}
We construct initial data satisfying the constraint equations
by numerically solving the CTS equations, discretized with 
second-order accurate finite differences, using a multigrid algorithm.
Details can be found in~\cite{East:2012zn}.
We then evolve the Einstein equations in the generalized harmonic formulation using the code 
described in \cite{Pretorius:2004jg,East:2011aa}, and assuming periodic boundary
conditions. See appendix~\ref{app:numerics} for more details on our numerical
methods.
In order to characterize our results we make use of the following diagnostic quantities.
We compute the total and individual energy densities defined according to \eqref{eq:energy_density}. 
We define a fiducial local Hubble expansion rate 
\begin{equation}\label{eq:hubble}
	H_{K} \equiv -\frac{K}{3}
\end{equation}
which allows us to compute the corresponding local number of $e$-folds of expansion 
$\mathcal{N}$ as 
\begin{equation}\label{eq:efolds}
	\mathcal{N} \equiv \int H_{K} d\tau \ ,
\end{equation}
where the integration is along the integral curve of $n^{\mu}$, and $\tau$ is the proper time
given by the lapse through $d\tau = N dt$ 
(see \cite{Xue:2013bva,East:2015ggf}). Note that in a homogeneous (i.e. FLRW)
spacetime this quantity is related to the initial and final scale factor as
$\mathcal{N}  = \int H d\tau = \log (a/a_0)$. 

The average of a quantity over the spatial volume is denoted by 
\begin{equation}\label{eq:average}
	\langle X \rangle \equiv \frac{\int d^3x \ \sqrt{\gamma} \ X}{ \int d^3x \ \sqrt{\gamma}}
\end{equation}
where $\gamma=\mathrm{det} \ \gamma_{ij}$.

In some cases (in particular, for smaller values of $k/H_0$), there is prompt gravitational
collapse in some regions and black holes form. To characterize the boundaries
of black holes in our dynamical setting, we use the concept of an apparent
horizon, defined as the outermost marginally outer trapped surface.  We 
excise a region interior to the black hole from the computational
domain, and do not evolve the equations there (see appendix~\ref{app:numerics} for more details on the
implementation). From the area of the apparent horizon $\mathcal{A}$ we
define the irreducible mass, $M_{\rm irr}=\sqrt{\mathcal{A}/16 \pi}$. 
We find that the angular momentum of the black holes is negligible, and hence the
irreducible mass closely approximates the Christodoulou mass.
To make
contact with observations, we compute the power spectrum of scalar fluctuations,
which restoring Planckian units and assuming slow-roll inflation is given by
\begin{equation}
	{\Delta_{\mathcal{R}}}^2 = \frac{1}{8\pi^2}\frac{H^2}{{M_P}^2}\frac{1}{\epsilon_{\scaleto{V}{3.5pt}}} \ ,
\end{equation}
evaluated at the time when the cosmic microwave background fluctuations crossed the horizon.
We also characterize the parameters we choose for the background cosmological solutions
(to which we add large inhomogeneities) by quoting the scalar spectral index and the tensor-to-scalar ratio, which
in the slow-roll approximation are given by
\begin{equation}
	n_s - 1 = 2 \eta_{\scaleto{V}{3.5pt}}  - 6 \epsilon_{\scaleto{V}{3.5pt}} , \ \ \ r = 
	16 \epsilon_{\scaleto{V}{3.5pt}} \ .
\end{equation}
\section{Results}\label{sec:results}
We examine the conditions under which an initially expanding universe, with either a single
scalar field, or two non-interacting scalar fields
(see section~\ref{sec:theory}), transitions to exponential expansion.
We construct initial data using the approach described 
in sections~\ref{sec:metric}-\ref{sec:matter}. 

We study two regimes of interest, distinguished by the energy scale at which inflation
takes place. 
We first consider single-field inflationary models 
where the universe is initially dominated by the gradient and kinetic energy of scalar field,
$\eta^2 \sim  \gamma^{ij} \partial_i \phi \partial_j \phi \sim 10^3 V(\phi_0) $, 
in which case the initial expansion rate is large compared to the inflationary Hubble rate
(we refer to this as ``low energy" or sub-Planckian inflation, see section \ref{sec:sub_planck}). 
We find that one or more (depending on the symmetry of the total energy density on 
the initial time slice)
regions of the domain undergo gravitational collapse, leading to the formation of black
holes. Ignoring the interior of the black hole(s), the spacetime is still expanding on
average, and we find that, independently of the type of initial data, the gradient and kinetic
energy dilute away until the numerical domain becomes dominated by inflationary energy.

We also explore inflationary models where the amount of kinetic, gradient, and
potential energy are all comparable on the initial time slice, and their sum is
order unity in units of the wavelength of the initial homogeneities (which is
also comparable to the initial Hubble radius) $\rho L^2 \lesssim O(1)$ (see
section \ref{sec:planck}).  Here, we refer to this scenario as high energy, or
nearly Planckian energy inflation, though, of course, our analysis is entirely
classical and ignores quantum effects.  We first consider single-field
inflationary models, where we find that the solution quickly transitions to
exponential expansion. Independently of the way initial data is constructed,
the late-time evolution always consists of an exponentially expanding universe
with a rapidly decreasing gradient energy and kinetic energy that asymptotes to
its corresponding slow-roll value.  We then extend our study to a two-field
inflationary model where the energy content is such that the first stage of
inflation may start at nearly Planckian energies, while the second phase will
occur at sub-Planckian energies. For technical reasons discussed below, we are
not able to evolve this scenario past the first stage of inflation. However,
already at this stage we find large patches where gradients have become
negligible and the spacetime is well described by a homogeneous evolution set
by the local inflaton values.  Extrapolating thusly indicates that these
regions should undergo a prolonged period of inflation driven by the second
field.

In the following, we quantify our results using the methods described in
section~\ref{sec:numerical_method}, by computing, for instance, the ratio of
kinetic and gradient energy to the potential energy.  In all the cases we
study, we consider perturbations such that the ratio of the wavenumber of the
initial scalar field content to the expansion rate on the initial time slice
$k/H_0$ is of the order of one, meaning our simulations probe the strong-field
regime.  As here we focus on models of large-field inflation, where $\phi_0$ is
large in Planckian units, and consider fluctuations of the the inflaton that
are of order one, we note that we do not consider scenarios where the
perturbations around the average scalar field value that exceed the distance in
field space to the end of inflationary plateau. Previous
studies~\cite{East:2015ggf,Clough:2016ymm,Aurrekoetxea:2019fhr} have found that
inflation can fail to happen in such scenarios. 

\subsection{Low energy inflation}\label{sec:sub_planck}
\begin{figure}[h]
\centering
\includegraphics[width=0.98\textwidth]{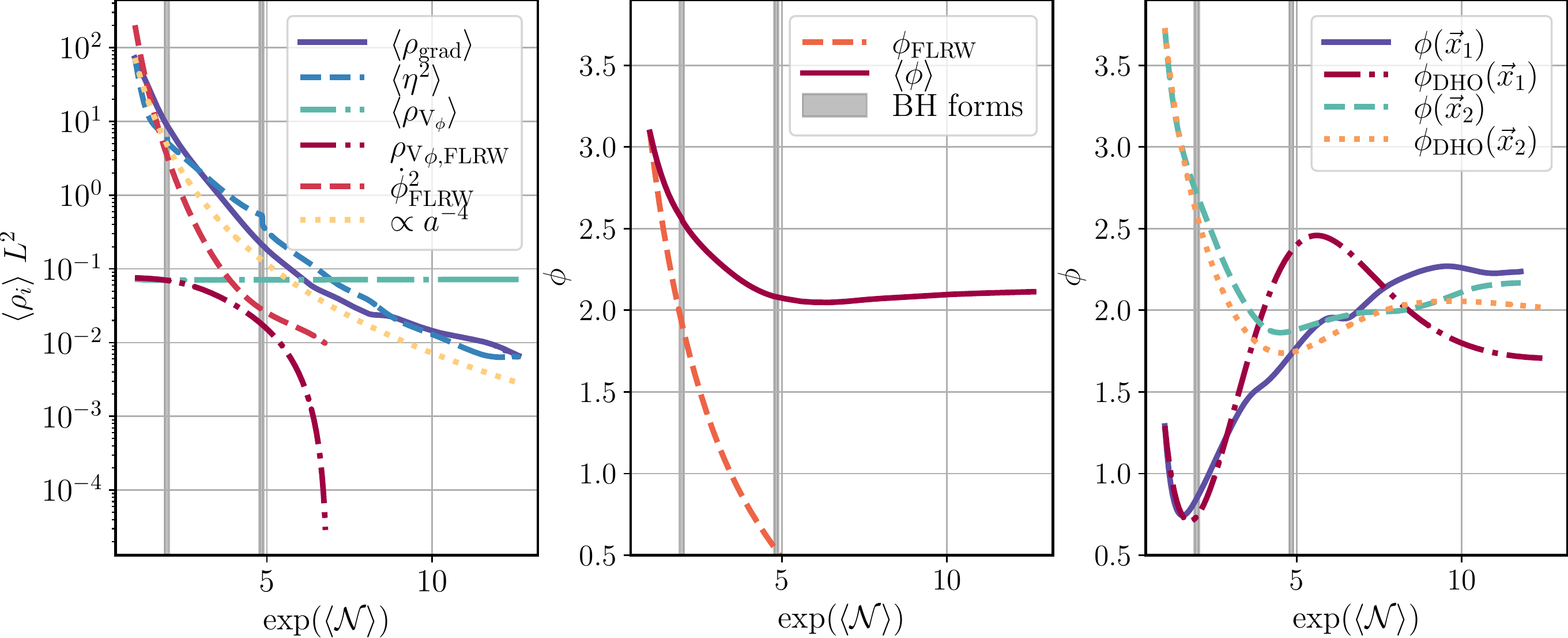}
	\caption{\label{fig:rho_NOPh} \emph{Left}:
	The volume-averaged kinetic $\eta^2$, 
	gradient $\rho_{\rm grad}$, and potential $\rho_{V_{\phi}}$ energy densities
	plotted against the averaged measure of the effective scale
	factor for the case where the time derivative of the scalar field is
	uniform on the initial time slice and the potential has a T-shape 
	\eqref{eq:NOTCH} with 
	$(\phi_0,m,\alpha,\omega,\delta \phi)=(3.0916,0.14,1/6,-7.1,0.9)$.
	For comparison, we also show
	the evolution of the kinetic energy density $\dot{\phi}_{\rm FLRW}^2$, 
	and potential energy density $\rho_{V,{\rm FLRW}}$ when solving the corresponding
	homogeneous FLRW equations in the absence of gradient energy and by
	specifying the initial value of $\dot{\phi}_{\rm FLRW}$ to be the time derivative
	of the scalar field on the initial time slice.
	The energy density for a radiation dominated universe is shown by
	the dotted yellow line.
	\emph{Middle}: 
	The volume average of inflaton field value in comparison to the homogeneous solution.
	\emph{Right}:
	The value of scalar field at the locations
	$\vec{x}_1=L(0,-1/4,1/4)$ and $\vec{x}_2=L(1/8,-1/4,1/4)$
	when solving full equations of motion, and when solving
	equation for damped harmonic oscillator \eqref{eq:SHO}.
	}
\end{figure}

We first consider solutions where the potential energy density is initially subdominant.
To begin with, we describe an illustrative case where the time derivative of the scalar field 
is uniform on the initial time
slice, i.e. given by eq.~\eqref{eq:PhistDotCnst}. 
We choose the potential to be given by an $\alpha$-attractor
potential eq.~\eqref{eq:NOTCH} with $\phi_0 = 3.1$,
such that, in the absence of inhomogeneities ($\delta \phi=0$), and if the time
derivative of the scalar field were zero,
the combination of parameters $(\phi_0,m,\alpha)$ would result in 60 $e$-folds of inflation, 
a scalar power spectrum of $\Delta_{\mathcal R}^2 \sim O(1)$, a scalar spectral index of
$n_s = 0.97$, and a tensor-to-scalar ratio of $r = 0.001$ 
for modes that cross the horizon 60 $e$-folds before the end of inflation.
However, we pick an initial value of $\partial_t \phi$ such that, in a homogeneous
Universe, the scalar field would evolve off the plateau of the inflationary potential 
without inflation occurring. 
The gradient and kinetic energy are chosen to be comparable
on the initial time slice, but 800 times larger than the potential energy density.
Figure~\ref{fig:rho_NOPh} shows the individual volume-averaged 
energy densities (left panel) as a function of the effective scale factor 
$a = \mathrm{exp}(\langle \mathcal{N}\rangle)$. We also plot the
kinetic $\dot{\phi}_{\rm FLRW}^2$ and potential energy $\rho_{\rm V_{\phi,\rm FLRW}}$
densities when solving the homogeneous FLRW equations in the absence of gradient energy
and specifying the initial value of $\dot{\phi}_{\rm FLRW}$ to be the time derivative
of the scalar field on the initial time slice given by eq.~\eqref{eq:PhistDotCnst}.

We find that in some regions, the maximum energy density quickly increases and
black holes form (at the times indicated by the grey shaded areas in
figure~\ref{fig:rho_NOPh}).  As was pointed out in previous studies
\cite{East:2015ggf,Clough:2016ymm}, the formation of black holes can be 
motivated using the hoop conjecture~\cite{thorne_hoop}. 
Indeed, the hoop conjecture predicts
that, if the mass of an overdensity exceeds the mass
of a black hole of the same size, then the overdensity will collapse to a black hole.
Following this argument, one can expect black holes to form when 
$\frac{4}{3}\pi k^{-3} \rho \sim k^{-1}/2$ which is equivalent to
$k/H_0 \sim 1$, or $\delta \phi \sim 1$ in Planck units.
Here, since the total energy density on the initial time slice is symmetric with respect to
positive and negative values of the scalar field, 
two identical black holes form in the periodic domain.
These black holes only create locally collapsing regions, and, after removing their interior
from the domain of integration, the spacetime is
still expanding on average. The gradient and kinetic energy are diluted until the 
spacetime is dominated by the potential energy, after which inflation starts.
As the effective scale factor of the spacetime increases, the proper distance between
the black holes also increases, as their density is also diluted by inflation.

Surprisingly, we find that the addition of gradient energy
allows inflation to occur in cases where it would not in the equivalent
homogeneous case. We will now demonstrate that 
the reasons are two-fold.
On the one hand, there is some resistance from the gradient pressure,
which acts as a restoring force, and hence
tends to pull the scalar field up the potential and away from the minimum.
On the other hand, the addition of gradient energy results in an increase in
the initial expansion rate, as a result of which the oscillations of
the inflaton are damped. We now confirm this picture,
and propose
a simple toy-model which allows us to reproduce our results and gain
some intuition.
Let us consider the Klein-Gordon equation
on the background of an FLRW spacetime,
\begin{equation}\label{eq:KG}
	\ddot{\phi} + 3 H \dot{\phi} -\gamma^{ij} \partial_i \partial_j \phi + 
	V_{\phi}' = 0
\end{equation}
where we have included the gradient terms in addition to the usual friction term,
and the \emph{dot} here denotes the time derivative with respect to the proper time $\tau$.
We further assume that the gradient energy 
can be approximated by a homogeneous
energy density initially equal to the volume average, and that scales with an inverse power $p$ of the scale factor
\begin{equation}\label{eq:rho_rad}
    \rho_{\mathrm{grad}} \approx  2 V_0 r (a/a_0)^{-p} \ ,
\end{equation}
where $V_0 \equiv V_{\phi}(\phi=\phi_0)$ is the initial potential in the absence of inhomogeneities,
and $r$ gives the ratio of the initial value of gradient energy density to the inflationary
energy density 	$r \equiv \frac{\langle \rho_{\rm{grad, 0}} \rangle}{2 V_0}$.
In addition, we assume that the gradient energy redshifts in the same way as
radiation, i.e. $p=4$.
In order to treat the spatial derivative term, we make the simplifying approximation that
$\partial^2 \phi \sim (ik/a)^2 \Delta$, 
where $a$ is the spatially homogeneous scale factor, and
we have introduced the deviation from the background solution
$\Delta \equiv \phi(t,x,y,z) - \chi_0(t)$, where $\chi_0(t)$ is the solution to
the equation of motion \eqref{eq:KG} assuming the third term vanishes.
Writing this in terms of the homogeneous number of $e$-folds, $d\mathcal{N} = H d\tau$, 
the equation of motion \eqref{eq:KG} then becomes
\begin{equation}\label{eq:SHO}
    H^2\frac{d^2\phi}{d \mathcal{N}^2} + A \frac{d \phi}{d\mathcal{N}} + B \Delta = -  V_{\phi}'
\end{equation}
where, since $H^2 = \rho/3$ with $\rho$ given by \eqref{eq:energy_density},
using \eqref{eq:rho_rad}, and further assuming
$\eta^2 \sim \rho_{\rm grad}$, the coefficients are
\begin{eqnarray}
	H^2 = {H}_V^2 \left(\frac{V}{V_0}+2 r e^{-p \mathcal{N}}\right),
\end{eqnarray}
and
\begin{eqnarray}
A = 3{H}_V^2 \left[\frac{V}{V_0}+2 r e^{-p \mathcal{N}}\left(1+\frac{p}{6}\right)\right],
\quad
    B = \frac{k^2}{e^{2 \mathcal{N}}}
\end{eqnarray}
where $H_{V}^2=2V_0/3$ is the initial inflationary Hubble rate.
This expression takes the form of a damped harmonic oscillator,
where $H^2$ is analogous to the mass, $A$ represents the Hubble friction, 
$B$ the restoring pullback force, and the potential gradient the driving
force pushing the inflaton down the potential. 
It is now clear that, for $r>0$, the additional gradient energy
not only increases the initial friction as well as the \emph{mass} term,
but also provides an additional pullback force for points away
from the average value of the inflaton. Note that, as the universe
expands, the contribution from the gradient energy 
to the friction and restoring terms is exponentially suppressed,
such that if the scalar field remains within the inflationary
part of the potential during the initial dynamical phase,
then it can transition into slow-roll inflation.

In the right panel of figure~\ref{fig:rho_NOPh}, we compare the results of
numerically integrating the damped harmonic equation \eqref{eq:SHO} to the
behavior of the full simulations at two arbitrarily located points.  We find
that the results are qualitatively the same and conclude that this simple
toy-model can be used to build intuition about the dynamics of the full
solution.  Similarly, the left panel shows that the gradient and kinetic
energy of the full evolution decreases as $a^{-4}$, as would be expected for a
radiation dominated universe.  We also note that the arguments given above 
suggest that other significant contributions to the expansion rate from inhomogeneities,
for example from a gravitational wave energy density, as studied
in~\cite{Clough:2016ymm}, should also mitigate the disruptive effect of a small
nonzero initial scalar field velocity.

\begin{figure}[h]
\centering
\includegraphics[width=0.98\textwidth]{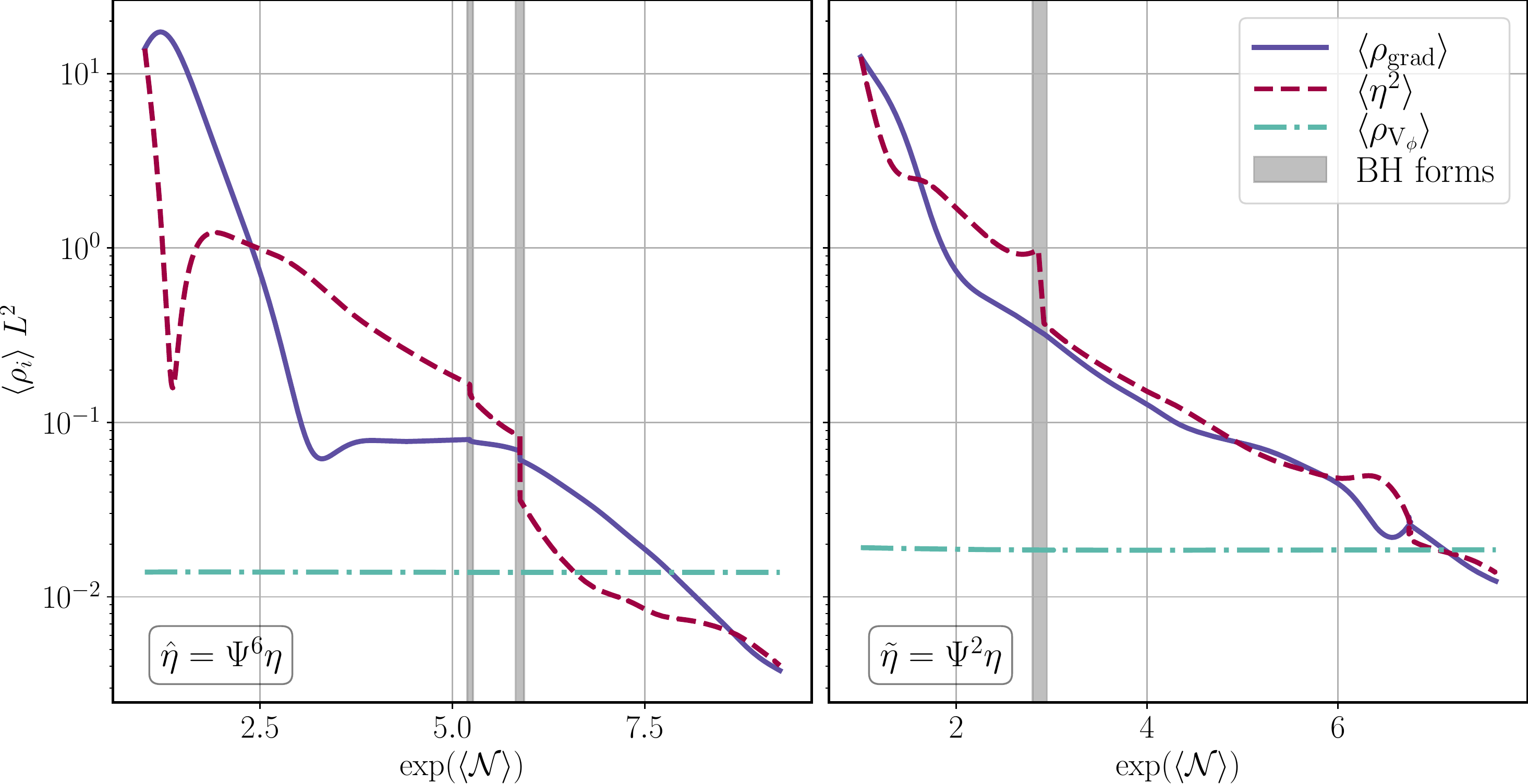}
	\caption{\label{fig:rho_Aij_NORF} 
	We show the gradient, kinetic, and potential energy
	contributions to the average energy density (similar to the left 
	panel of figure \ref{fig:rho_NOPh}) for two different cases. Gray
	regions indicate the formation of black holes.
        \emph{Left}:
	A case with a quadratic potential given by \eqref{eq:Vm} and
	initial data that automatically satisfies momentum constraint
	\eqref{eq:mom_triv} with parameters $(\hat{\phi}_0,m,\sigma,\delta \hat{\phi},\delta
	\hat{\eta})=(\sqrt{2} 11.0,3.78 \times 10^{-3},0.1,0.68,2.21)$ 
	\emph{Right}:
	A case with initial
	data constructed according to \eqref{eq:RF_stkin} 
	and with parameters 
	$(\phi_0,m,\alpha,\delta \phi)=(5.42,0.029,1.0,0.65)$ for the $\alpha$-attractor potential
	\eqref{eq:NOTCH}.
	 }
\end{figure}

We find that qualitatively similar results hold
when starting from initial conditions constructed 
by specifying a rescaled version of 
the conjugate momentum such that $\tilde{\eta}= \Psi^2 \eta$ \eqref{eq:RF_stkin}, 
or such that
the momentum constraint constraint is analytically satisfied $\hat{\eta}= \Psi^6 \eta$
\eqref{eq:mom_triv}.
In both cases, the rescaled conjugate momentum is chosen such that the gradient and kinetic
energy are comparable
on the initial time slice, but $\sim 1000$ times 
larger than the potential energy density.
The former is demonstrated in the
right panel of figure~\ref{fig:rho_Aij_NORF}, where we choose the potential given
by an $\alpha$-attractor potential \eqref{eq:NOTCH}, and an initial scalar 
field profile given by \eqref{eq:RF_stkin}, with parameters
such that, in the absence of inhomogeneities ($\delta \phi=0$), and if the conjugate momentum
of the scalar field were zero,
the combination of parameters $(\phi_0,m,\alpha)$ 
would result in 60 $e$-folds of inflation, 
a scalar power spectrum of $\Delta_{\mathcal R}^2 \sim 0.05$, a scalar spectral index of
$n_s = 0.97$, and tensor-to-scalar ratio of $r = 0.006$ 
for modes that cross the horizon 60 $e$-folds before the end of inflation.
We find that the gradient and kinetic energy decrease until the spacetime is dominated
by the inflationary energy after which inflation may start. Unlike for initial data with
a constant velocity profile, here the energy density on the initial time slice is no longer
symmetric with respect to positive and negative values of the scalar field, and therefore
the black holes are no longer symmetric about the origin.
Similarly, the left panel of figure~\ref{fig:rho_Aij_NORF} 
shows the individual energy densities for a solution constructed according to \eqref{eq:mom_triv} 
and with quadratic
potential such that in the absence of inhomogeneities and conjugate momentum,
the combination of parameters $(\phi_0,m)$ would result in 60 $e$-folds of inflation,
$\Delta_{\mathcal R}^2 \sim 1\times 10^{-3}$, $n_s = 0.97$, and $r = 0.13$
for modes that cross the horizon 60 $e$-folds before the end of inflation.
Here, again, two regions collapse to black
holes, but the universe keeps expanding, and eventually transitions to exponential expansion.
These results emphasize that our conclusions are generic, and not sensitive 
to the particulars of how the scalar field momentum is chosen.

\subsection{High energy inflation}\label{sec:planck}
\subsubsection{Single-field inflation}
\begin{figure}[h]
\centering
\includegraphics[width=0.98\textwidth]{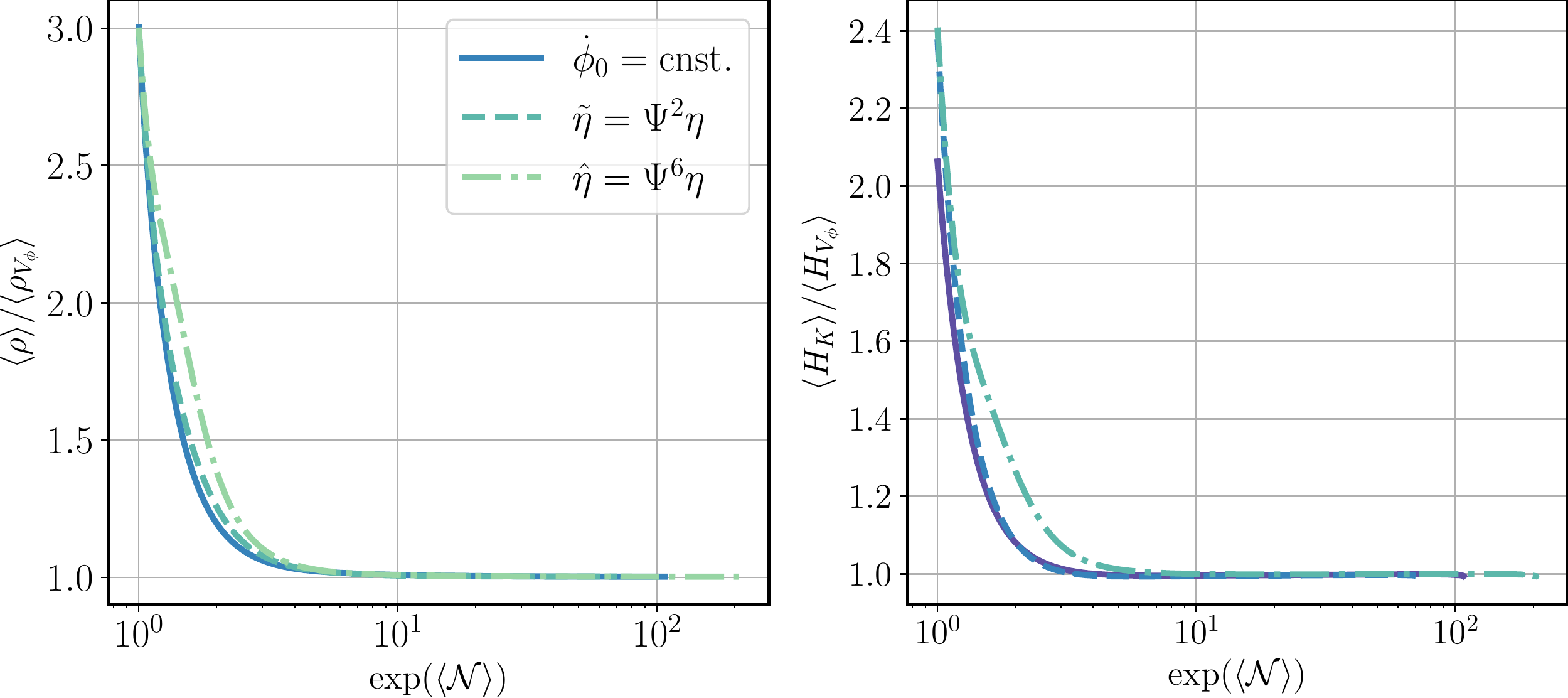}
\caption{\label{fig:rhos_planck} The volume-averaged total energy density (left) and 
	expansion rate (right) plotted against a volume-averaged measure of the effective scale
	factor for cases with a quadratic potential given by \eqref{eq:Vm}. 
	The quantities $\langle \rho_{V,\phi} \rangle$ and $\langle H_{V_{\phi}}\rangle$ respectively represent
	the volume-averaged measures of the potential energy density and inflationary
	expansion rate as a function of time.
        Both quantities can be seen to asymptote to the values expected for potential energy 
	dominated inflation. 
	The solid 
	line represents a solution for initial data constructed using \eqref{eq:PhistDotCnst}
	and $(\phi_0,m,\omega,\delta \phi)=(11.0,0.23,-5.83,0.79)$. The
	dashed line represents initial data given by \eqref{eq:RF_eta} with 
	$(\phi_0,m,\delta \phi)=(11.0,0.13,0.85)$.
	Finally, the dash-dotted curve represents initial data given by \eqref{eq:eta_trivial}
	and $(\hat{\phi}_0,m,\sigma,\delta \hat{\phi},\delta \hat{\eta})=
	(\sqrt{2} \ 11.0,0.12,0.1,0.68,2.21)$. 
	}
\end{figure}

\begin{figure}[h]
\centering
\includegraphics[width=0.98\textwidth]{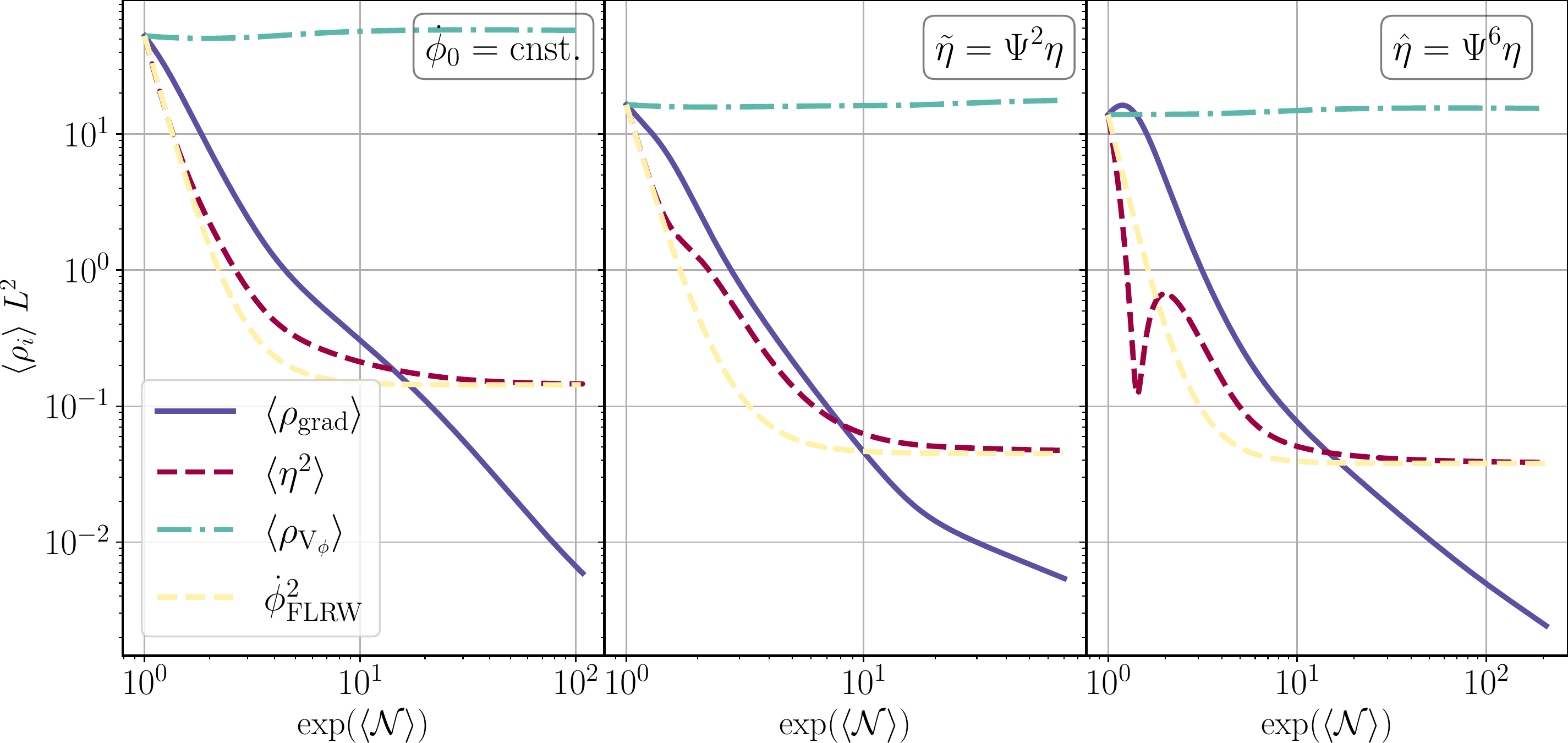}
	\caption{\label{fig:rhoi_planck} The volume-averaged kinetic $\eta^2$, 
	gradient $\rho_{\rm grad}$ and potential $\rho_{V_{\phi}}$ energy densities 
	plotted against the averaged measure of the effective scale
	factor for the cases shown in figure~\ref{fig:rhos_planck}.
	For comparison, the evolution of the kinetic energy density when
	solving the FLRW equations in the absence of gradient energy
	and specifying the initial value of $\dot{\phi}_{\rm FLRW}^2$ to be
	the volume average value of the kinetic energy of inhomogeneous
	solution on the initial time slice,
	is also shown (dashed yellow line).
        }
\end{figure}

We next consider single-field inflationary models where inflation may naturally begin 
near the Planck scale. Thus, solutions where the sum of the kinetic, gradient and
potential energy of the inflaton is order unity in Planck units.
We further restrict our study to initial conditions where
the kinetic, gradient, and potential energy are initially
comparable, and the potential is quadratic in the scalar field, $V_{\phi}(\phi)= m^2 \phi^2$. 
In figure~\ref{fig:rhos_planck}, we show the volume-averaged total energy density (left panel) 
and the volume-averaged expansion rate (right panel) 
computed from \eqref{eq:hubble}, as a function of the effective scale factor,
for the three types of initial data described in section~\ref{sec:matter}.
We specify the initial data 
such that, in the absence of gradient and kinetic energy, the homogeneous
initial value of $\phi_0 = 11$ and other
parameters would result in 60 $e$-folds of inflation,
$\Delta_{\mathcal R}^2 \sim O(1)$, $n_s = 0.97$, and 
$r = 0.13$ for modes that cross the horizon 60 $e$-folds
before the end of inflation.
Independent of the type of initial data used, as the scale factor increases, the average 
expansion rate and total energy density decreases (although at different rates
depending on the initial data), 
until the kinetic and gradient energy
is subdominant, leading to a universe dominated by the inflationary potential, 
hence undergoing exponential expansion.
This is illustrated in figure~\ref{fig:rhoi_planck}, where we show the individual average
energy densities. Although the gradient energy quickly becomes negligible, the kinetic
energy asymptotes to a constant value, consistent
with the value the kinetic energy would asymptote to
when solving the homogeneous FLRW equations in the absence of gradient energy
and specifying the initial value of $\dot{\phi}_{\rm FLRW}^2$ to be
the volume average value of the kinetic energy of inhomogeneous solution
on the initial time slice.

Though we evolve for long enough that the spacetime is dominated by the inflation potential
energy and exponential expansion, we do not evolve to the end of inflation, as 
growing gradients from different regions in the domain inflating at different rates
eventually lead to large truncation error.

We confirm that inhomogeneities do not prevent inflation in models
of chaotic inflation occurring near the Planck scale 
\cite{Linde:1984ir,Linde:1985ub,Linde:2014nna}.
Although the simplest models of inflation (including the
quadratic potential we study here), are now strongly disfavored by the most
recent observations from BICEP/Keck and Planck \cite{Planck:2018jri,BICEP:2021xfz},
they could still appear in the context of more complicated inflationary models
containing more than one scalar field. We discuss an example of such model in the following
section.

\subsubsection{Two-field inflation}\label{sec:two_fields}
Up to this point, we have only considered single field models. However,
our methods can be applied equally well to more complicated inflationary
models. To give an example, we consider a simple theory describing two
non-interacting scalar fields, $\theta$ and $\phi$, such that the 
potential is given by \eqref{eq:2field}, which combines a quadratic
dependence for $\phi$ with an $\alpha$-attractor term for $\theta$.
We further assume that $M \gg m$. Here, the potential can give rise to two stages of
inflation. The first stage of inflation is driven by the quadratic potential, and
may start at nearly Planckian energies and be relatively short. Once inflation
driven by $\phi$ ends, the second stage of inflation driven by the $\alpha$-attractor 
potential, and lasting more than 60 $e$-folds, may begin.
Here we are not able to evolve this model past the first stage of inflation, likely
due to limited numerical resolution and the gauge choice. However, already at this stage
we find that gradient energy has become subdominant, and by comparing the evolution in different regions 
to the homogeneous evolution with the same scalar field values, we can approximately extrapolate their later behavior.

\begin{figure}[h]
\centering
\includegraphics[width=0.98\textwidth]{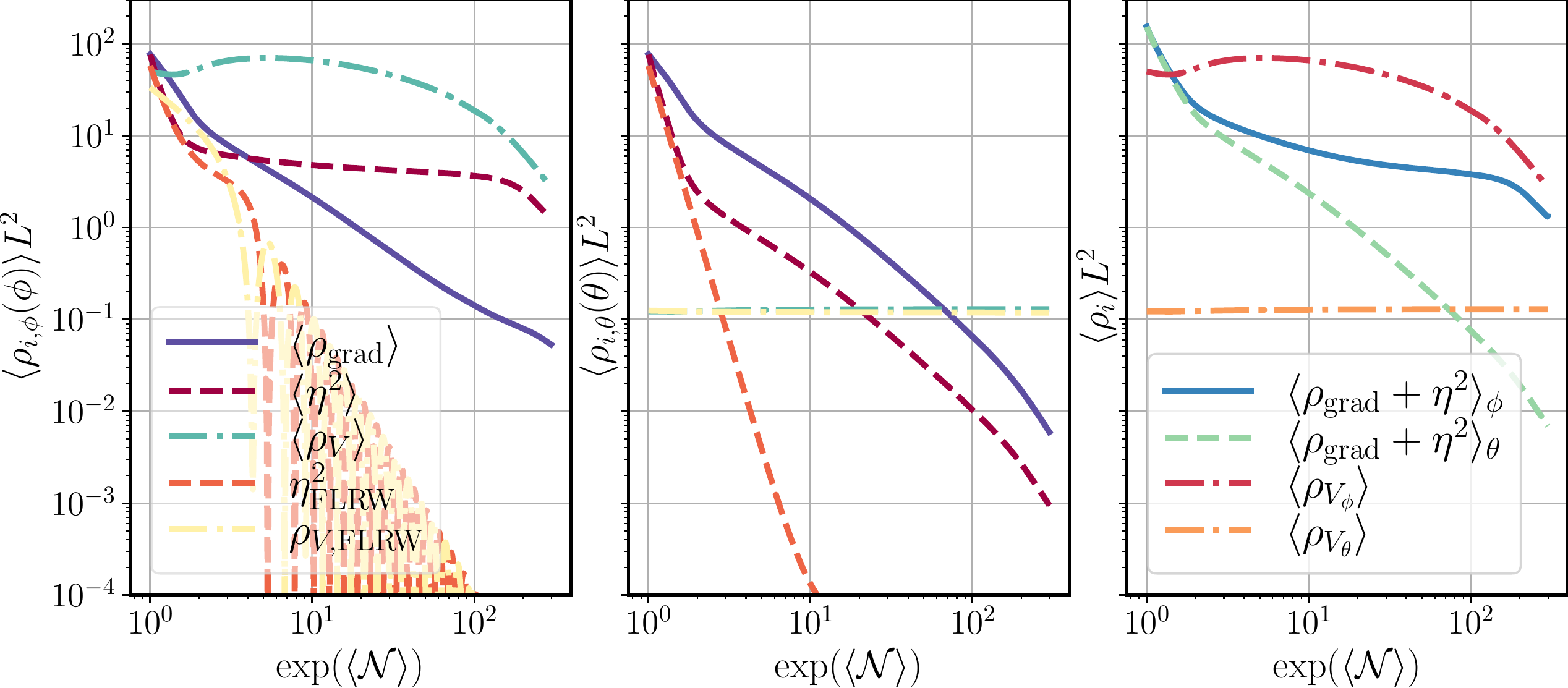}
        \caption{\label{fig:rhoi_two} 
	The gradient $\rho_{\rm grad}$, kinetic $\eta^2$, and potential energy
	$\rho_{V_{\phi}}$
        contributions to the volume average energy density as a function
	of the effective scale factor for each individual scalar field
	($\phi$ left panel and $\theta$ middle panel) and both scalar fields combined (right panel).
        \emph{Left}:
	Individual contributions to average energy density for scalar field
	with quadratic potential driving first
	stage of inflation. 
	For comparison, the evolution of the kinetic energy density $\eta^2_{\rm FLRW}$ 
	and potential energy density ${\rho}_{V,\rm FLRW}$ when
        solving the FLRW equations in the absence of gradient energy,
        and specifying the initial value for the time derivative of scalar field
	to be
        the volume average value of the conjugate momentum of inhomogeneous
	solution on the initial time slice, is also shown.
        \emph{Middle}:
	Same as the left panel, but for a scalar field with the $\alpha$-attractor
	potential driving second stage of inflation.
        \emph{Right}:
	The sum of the volume-averaged gradient and kinetic energy densities
	for each scalar field separately $\langle \rho_{\rm{grad}}+\eta^2 \rangle$ 
	and the volume-averaged potential energy densities for each scalar field.
        }
\end{figure}

We consider values for $\phi_0$ and $\theta_0$ such that, in the absence
of inhomogeneities ($\delta \phi=0$ and $\delta \theta=0$) and kinetic energy on
the initial time slice,
the first stage of inflation would last one $e$-fold, and 
the second stage of inflation would last 60 $e$-folds and give rise 
to a scalar power spectrum with 
$\Delta_{\mathcal R}^2 \sim 0.1$, $n_s = 0.97$, and $r = 0.006$.
The remaining free data is chosen such that 
the total energy density is initially on the order of $1/L^2$,
the potential energy is 1/7 of the total energy, and the remaining fraction is provided by the
kinetic and gradient energy of the fields $\theta$ and $\phi$, with each term,
for each field
contributing equally to the total energy content.
We choose an initially uniform time derivative of the scalar field
for ease of implementation,
but based on the previous section we do not expect our results to
depend on the initial velocity profile of the scalar fields.
The conjugate momentum on the initial time slice
is such that, if one were to solve the FLRW equations
in the absence of gradient energy,
but specifying the initial values for
$(\dot{\phi}_{\rm{FLRW}},\dot{\theta}_{\rm{FLRW}})$
to be the volume average of conjugate momenta of
the corresponding scalar fields on the initial time slice,
then the first stage
of inflation would only last half an $e$-fold and the second stage $35$.
The left panel of 
figure~\ref{fig:rhoi_two} shows the individual contributions to the total energy density
for each scalar field in our simulation compared to when solving
the FLRW equations in the
absence of gradient energy and by specifying the initial value for the time derivative 
of scalar field to be the volume average value of the conjugate momentum of inhomogeneous
solution on the initial time slice.

\begin{figure}[h]
\centering
\includegraphics[width=0.98\textwidth]{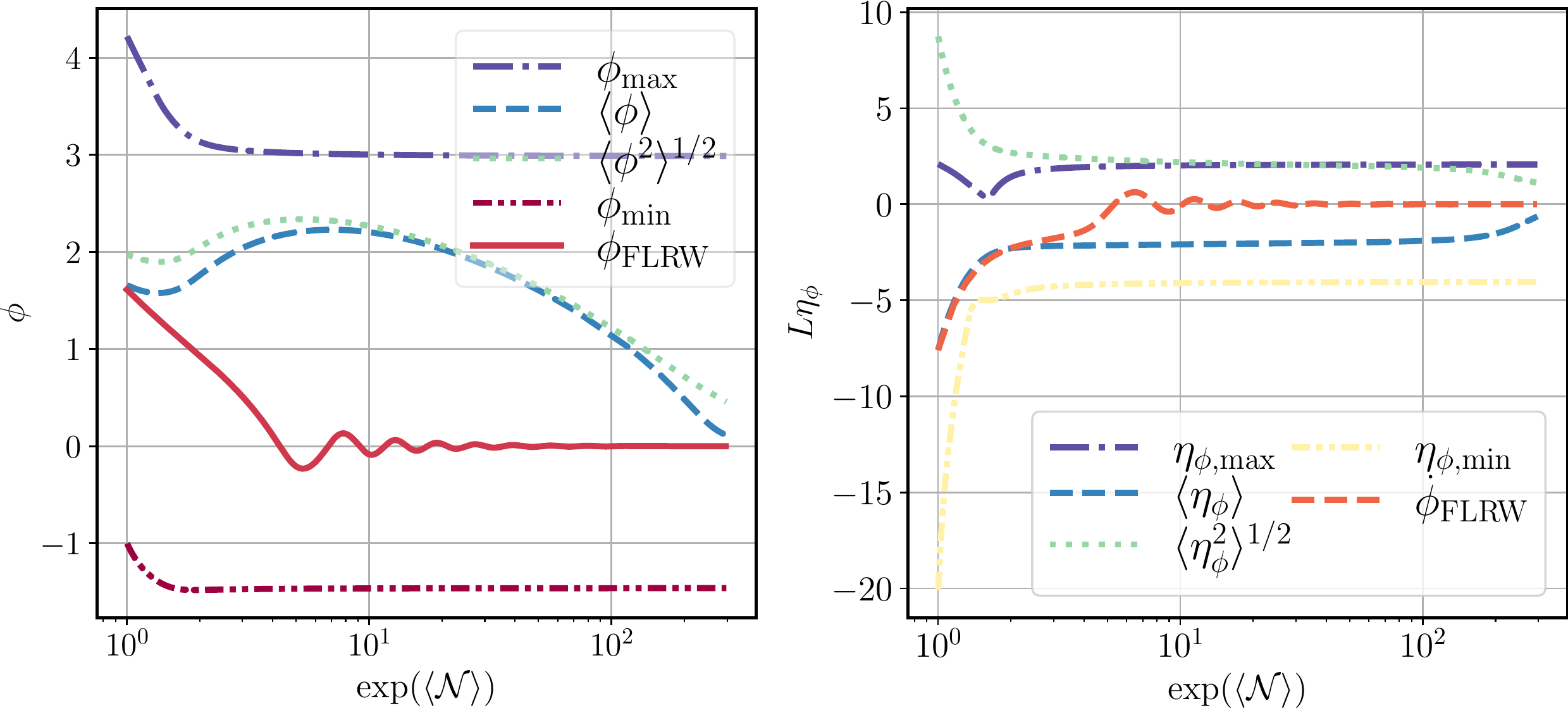}
        \caption{\label{fig:sf_two} 
	\emph{Left}: The minimum $\phi_{\rm{min}}$, maximum $\phi_{\rm{max}}$, 
	and volume-averaged values of the scalar $\langle \phi \rangle$ driving the
	first stage of inflation
	as a function of effective scale factor for the case shown 
	in figure~\ref{fig:rhoi_two}. 
	For comparison, we also show the evolution of $\phi_{\rm FLRW}$, the scalar field
        when solving the FLRW equations in the absence of gradient energy,
	and specifying the initial value for the time derivative of scalar field
        to be
        the volume average value of the conjugate momentum of inhomogeneous
        solution on the initial time slice.
	\emph{Right}: Similar quantities, but for the conjugate momentum of the scalar.
	}
\end{figure}

We first study the first stage of inflation driven by the quadratic potential,
shown in the left panel of figure~\ref{fig:rhoi_two}, in detail.
In line with our earlier observations, we find that the average kinetic energy of the scalar
field $\langle \eta^2_{\phi} \rangle $
(dashed red line) does not decrease as quickly as
in the homogeneous case (dashed orange line). We also
note that the average potential energy of $\phi$ (dash-dotted green line)
increases at first, as does the average scalar field $\langle \phi \rangle$ initially,
shown in figure~\ref{fig:sf_two},
likely pulled up by the gradients in the field. 
As the universe expands, the gradient energy
decreases and eventually, when the gradient energy is small enough, the potential
energy starts decreasing. 

Unfortunately, we were not able to able to evolve this scenario past the first 
stage of inflation.
We attribute this to a lack of resolution to resolve large gradients developing
between different regions of the domain inflating at different
rates. We next illustrate these different regions and conjecture that if we had
sufficiently large resolution, then the effectively causally
disconnected Hubble patches will keep inflating, albeit at different rates,
until the second stage of inflation kicks in.

Figure~\ref{fig:sf_two} shows the range of values the scalar field driving
the first stage of inflation and the conjugate momentum take, as a function of
the effective scale factor. This 
already suggests that different patches will inflate at different rates, which 
in turn will lead
to the formation of gradients in the field values across the numerical domain.

\begin{figure}[h]
\centering
\includegraphics[width=0.98\textwidth]{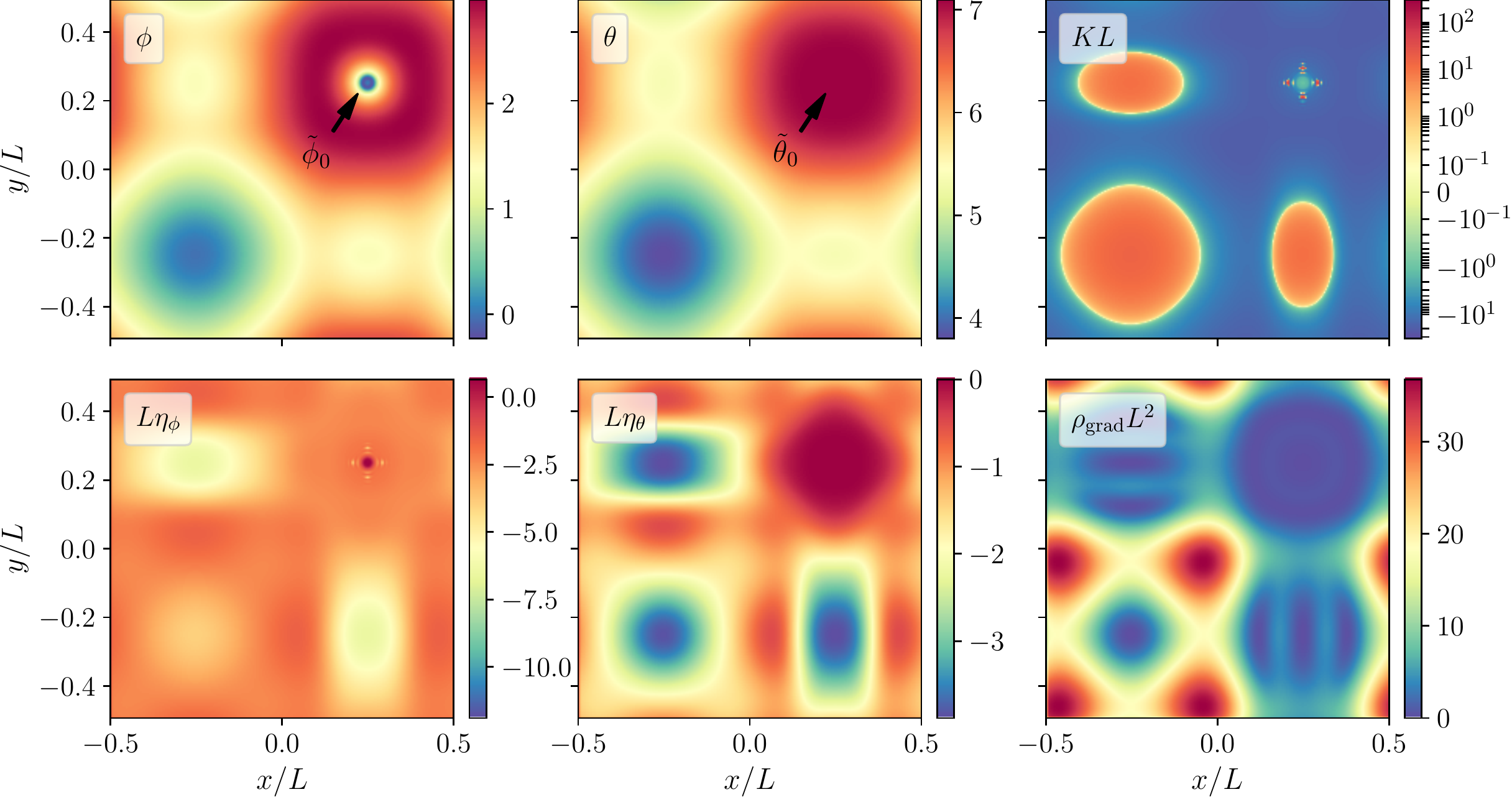}
        \caption{\label{fig:zNan} 
	Two-dimensional spatial slices showing 
	different quantities shortly before strong gradients (around the point indicated by the arrow)
        make the evolution inaccurate (at the time indicated by vertical
	dashed grey line in figure~\ref{fig:flrw_extrapol})
	for the case shown in figure~\ref{fig:rhoi_two}.
	The first column shows the scalar field (top) driving the first stage of inflation
	and its conjugate momentum (bottom). The second column shows the corresponding quantities
	for the scalar driving the second stage of inflation. The third column shows
	the extrinsic curvature scalar (top) and total gradient energy.
	    }
\end{figure}

In particular, the region where sharp gradients develop in the field values,
which we eventually can no longer resolve due to the finite numerical resolution
of our grid,
corresponds to 
where the scalar field $\phi$ stops inflating first. 
Figure~\ref{fig:zNan} shows the scalar
field $\phi$, its conjugate momentum $\eta_{\phi}$, the extrinsic curvature $K$,
and the gradient energy $\rho_{\rm{grad}}$ 
at the location where these unresolved features develop (indicated by $\tilde{\phi}_0$),
but shortly before these become severe enough that the evolution must be
halted (indicated by vertical dashed grey line
in figure~\ref{fig:flrw_extrapol}).
We find that the extrinsic curvature is larger than in the rest of the domain,
and that the gradient energy is
negligible, as expected.
We also show the value of the scalar field $\theta$ and its conjugate momentum $\eta_{\theta}$ 
for the field
driving the second stage of inflation. These quantities indicate that the
scalar field driving the second stage of inflation is still inflating at that location.

As the universe expands, the left and middle panels of figure~\ref{fig:rhoi_two} show that the average
gradient energies of $\phi$ and $\theta$ are quickly diluted until they are negligible compared 
to the average inflationary energy of either field. This suggests that when the gradient energy becomes
negligible in a neighbourhood surrounding a point of interest, then one can use the FLRW
equations together with the local value of scalar field and its time derivative to compute
the evolution of scalar field at that location. We first show that this approximation is valid 
for the location where unresolved gradients eventually develop. We then
use this to extrapolate the behavior of both scalar fields, at different locations,
past the point we are able to evolve.

In the left panel of figure~\ref{fig:flrw_extrapol}, 
we plot the values of the respective scalar fields at the location 
where unresolved 
features eventually develop $(\tilde{\phi}_0,\tilde{\theta}_0)$ as a function of the local
scale factor (purple dots). The smallest value of the scale factor for which we show 
the values of the scalar fields corresponds to the time at which the gradient
energy becomes negligible in a neighbourhood surrounding the point of interest.
We also show the evolution of the scalar field when solving 
the FLRW equations (dash-dotted lines),
ignoring the gradient terms, and fixing the initial values for the scalar field
and its time derivative to be the corresponding values for the inhomogeneous scalar
fields and conjugate momenta at that spatial point.
As expected, we find close agreement between the numerical and FLRW solution, and
make use of the FLRW
solution to extrapolate the behavior of scalar fields past the point
we were able to numerically evolve. For the spatial point where $\phi$ stops inflating
first, we find that $\theta$ will still inflate for $\sim 250$ $e$-folds 
at that particular point.

\begin{figure}[h]
\centering
\includegraphics[width=0.98\textwidth]{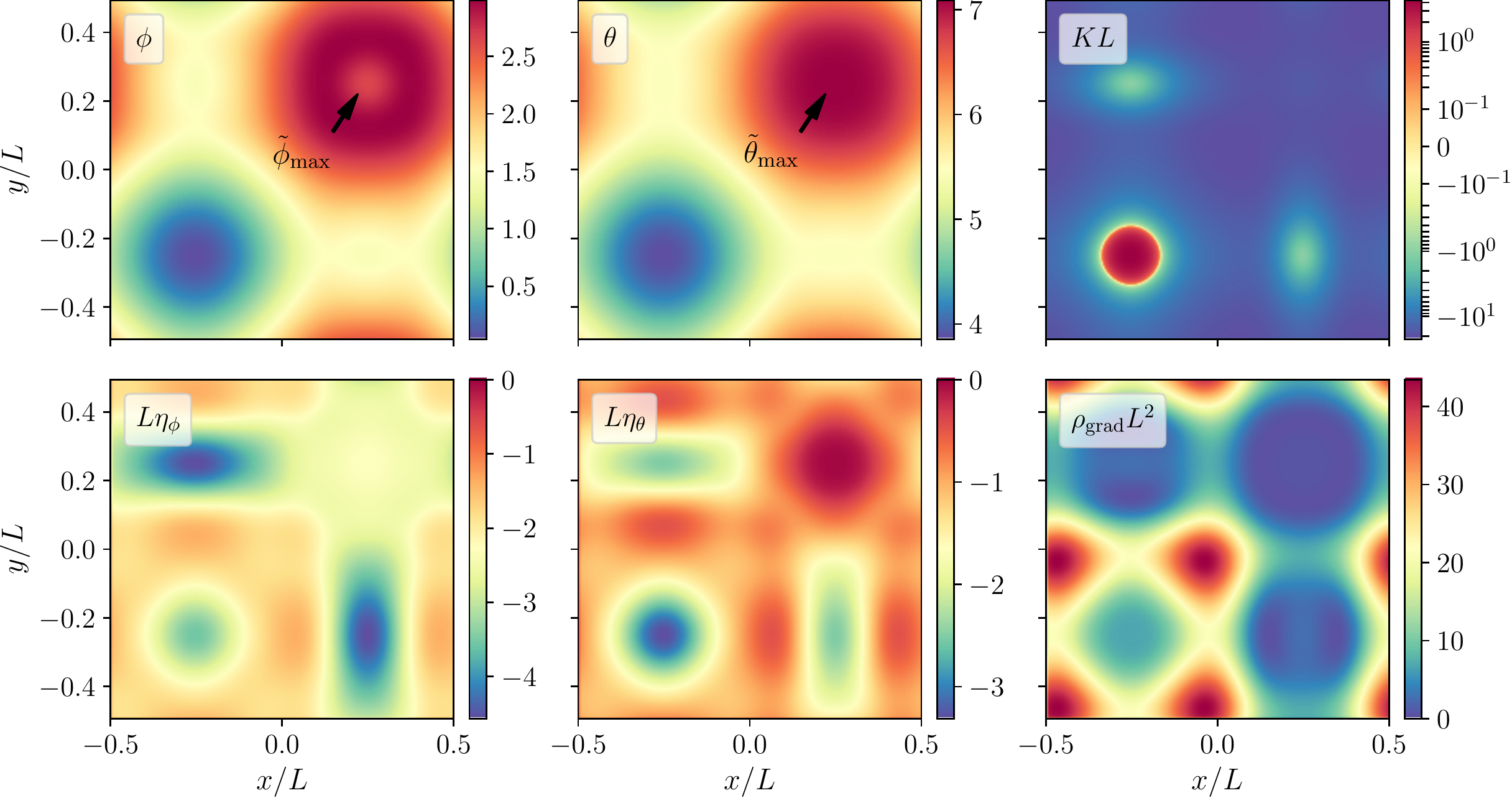}
        \caption{\label{fig:phi4_r_max} 
	Same as figure~\ref{fig:zNan}, but for the two-dimensional spatial slices
	where $\phi$ is maximal. }
\end{figure}

\begin{figure}[h]
\centering
\includegraphics[width=0.98\textwidth]{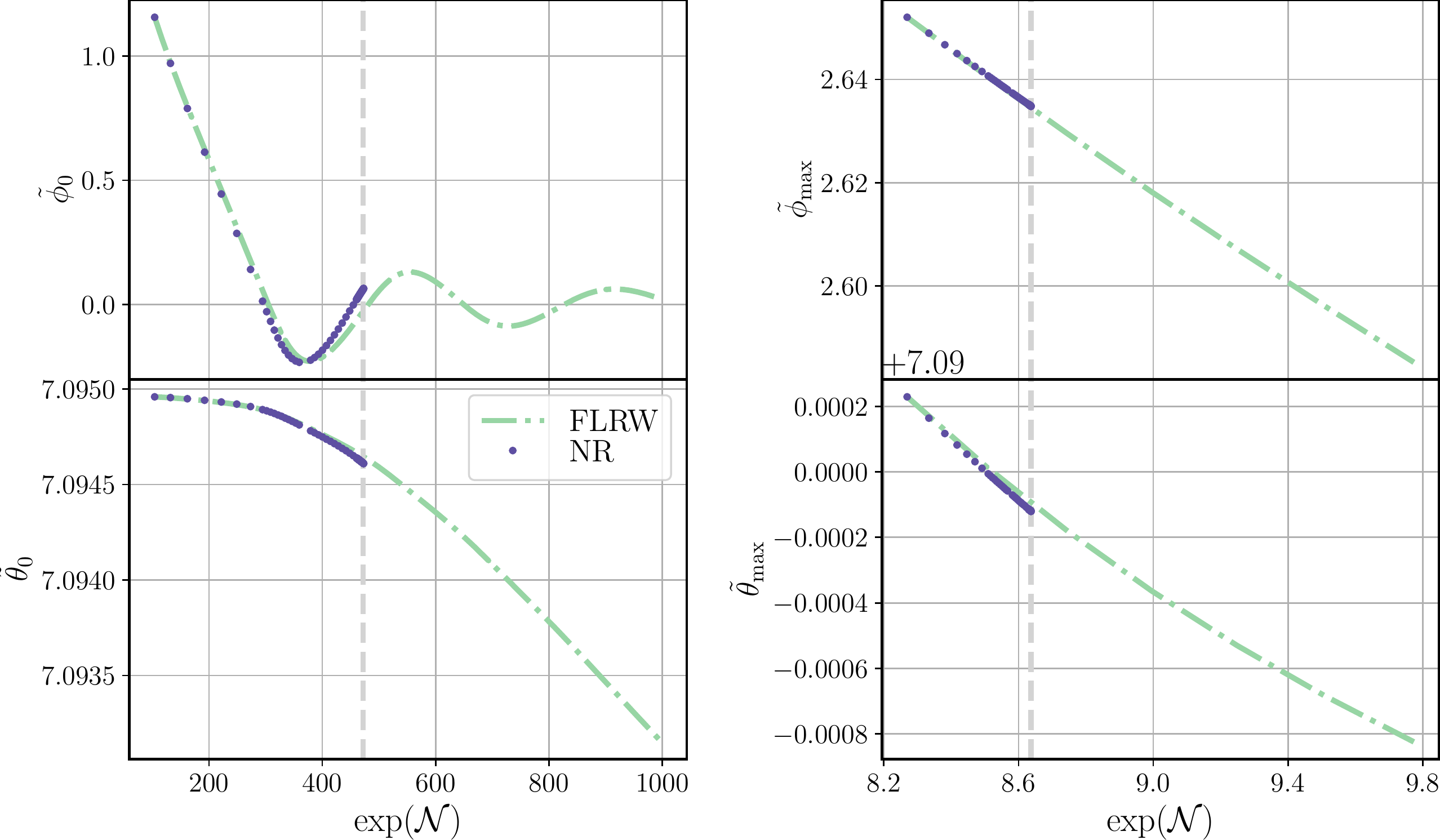}
        \caption{\label{fig:flrw_extrapol}
	Scalar field and conjugate momentum values as a function of the local
	effective scale factor for two spatial points (indicated by $\tilde{\phi}_0$ and
	$\tilde{\phi}_{\rm max}$ on the spatial slices in figures~\ref{fig:zNan}
	and \ref{fig:phi4_r_max}, respectively), chosen such that gradient energy is negligible in
	surrounding neighbourhood (blue dots).  The corresponding homogeneous FLRW
	solution using the same initial values plotted in the figure is also shown. 
	The vertical dashed grey lines indicate the local scale factors
	corresponding to the
	two-dimensional spatial slices shown in figures~\ref{fig:zNan} and ~\ref{fig:phi4_r_max}.
	}
\end{figure}

As another representative example,
we also show 
the scalar
field, its conjugate momentum, the extrinsic curvature,
and the gradient energy
at the location where $\phi$ obtains its maximum value (indicated by $\tilde{\phi}_{\rm{max}}$
in figure~\ref{fig:phi4_r_max}).
Similarly, we find that the gradient energy is negligible in a neighbourhood of that point and 
plot the scalar field values and their corresponding FLRW solutions as a function
of the local scale factor in the right panel of figure~\ref{fig:flrw_extrapol}.
We extrapolate the FLRW solution, and deduce that $\phi$ and $\theta$ will inflate for an additional
$3$ and $250$ $e$-folds respectively.

These results suggest
that other parts of the domain will also keep inflating,
although at different rates determined by the local value of the
scalar field and its conjugate momentum at that location.
We thus conjecture that, if we 
were able to continue the evolution, $\phi$
would fluctuate around zero in regions where it has
reached the bottom of the potential (just like in the homogeneous case).
We leave a more detailed analysis to confirm this to to future work.

Assuming the above picture, as 
the universe keeps expanding, the potential energy of the $\alpha$-attractor potential 
eventually dominates the quadratic potential, after which the second stage 
of inflation starts.
The middle panel of figure~\ref{fig:rhoi_two} shows that the gradient 
and kinetic energy of the scalar field driving the second stage of inflation
becomes smaller than the potential energy even before the $\alpha$-attractor potential 
dominates over the quadratic potential.
Furthermore, 
we find that (shortly before large gradients make the evolution inaccurate) 
the wavelength of the perturbations $\sim L$ is about $60$ times
larger than the volume averaged Hubble radius of scalar field driving second stage of inflation.
This suggests that by the time the first stage of inflation ends,
the universe is effectively made up of patches that are homogeneous on Hubble scales and,
at least in most regions, the second stage of inflation
may generate a nearly-scale invariant spectrum consistent with observations.

\section{Discussion and conclusion}\label{sec:conclusion}
In this paper, we have undertaken the first study of the robustness of large
field models of inflation to large initial gradients and kinetic terms in the
initial scalar field profile.  We proposed three different ways of specifying
scalar field configurations with non-vanishing momentum density when solving
the Hamiltonian and momentum constraint equations using the CTS formalism.
Solving the full system of coupled constraint equations allowed us to examine
the effects of several initial conditions not previously studied in the
literature. Our results are summarized as follows:

\begin{itemize}

\item We studied the evolution of single-field inflationary models where the initial
gradient and kinetic energy densities of the inflaton are much larger than
potential energy. We demonstrated that the presence of significant gradient
energy can actually negate the effect of a large time derivative in the scalar
field, allowing inflation to occur in cases where, in the homogeneous case,
the kinetic energy would drive the scalar field off the inflationary plateau
before exponential expansion occurred.  We explained this in terms of the
increased initial expansion rate and restoring pullback force in the presence
of large gradients, deriving a simple toy-model to demonstrate this.  This
implies that rather natural initial gradients in the initial conditions will
mitigate the effects of a non-zero initial field velocity of the inflaton. 

\item We also performed the first study of the impact of inhomogeneities on
single-field inflationary models where inflation may start at nearly Planckian
densities.  We found that, in cases where the gradient, kinetic and
potential energy densities are all comparable at nearly Planckian densities, the
universe will rapidly transition to exponential expansion. In these cases we did not
find black hole formation.

\item In addition to single-field inflationary models, we performed a
preliminary study of the effects of inhomogeneities on cosmological
scenarios where the universe undergoes two stages of inflation, the
first one at nearly Planckian energies and the second one at
sub-Planckian scales consistent with observations. Although we were not
able to evolve the spacetime to the point where the second stage of
inflation would start, and therefore make detailed conclusions, we find
at this time large regions where the gradients are negligible and the
spacetime is well approximated by the homogeneous evolution determined
by the local inflaton values.  We leave a more detailed study of these
models for future work.

\end{itemize}

Exploring the two-field inflationary models considered here
would require developing new numerical methods.  As
mentioned above, we believe one of the main difficulties lies in being able to
resolve large gradients developing between different regions of the numerical
domain inflating at different rates.  A better choice of gauge, as well as
working with higher resolution or adaptive mesh refinement may help address
this, as well as allow us to evolve the single-field
inflationary models for longer.

In this study, we only considered two families of inflationary potentials out
of the multitude that have proposed in the literature.  However, we verified
that, starting from the same initial conditions, these two different potentials
gave similar qualitative results and we do not expect our results to depend
strongly on the details of the potential, since the key features are the
flatness of the potential and the super-Planckian distance in field space
that must be traversed to reach the minimum of the potential. 
For two-field inflationary models, for simplicity in this first study, we
focused on theories where the scalar fields are non-interacting. It would be
interesting to consider the evolution of two-field inflationary models where
the scalar fields do interact and explore whether there are still robust to
generic initial conditions (see e.g. \cite{Linde:2014nna} and \cite{Joana:2022uwc}).
Note however, 
that in the $\alpha$-attractor generalization \cite{Kallosh:2022ggf,Braglia:2022phb}
of the hybrid inflation scenario \cite{Linde:1993cn}, 
the fields become decoupled in the large field limit, 
and the evolution of the initial conditions reduces to the one performed in our paper.

Another direction for future work would be to adapt the methods developed here
to study the robustness inflation in the presence of other types of matter.
One particular class of models of interest would be matter where the speed of
sound of the fluctuations, or equivalently the Jeans length, goes to zero
\cite{Senatore:2016aui}, as in, for example, dust.  This would be a case
especially prone to the gravitational collapse of initial perturbations.

The methods developed here for constructing general relativistic initial data
are rather generic, and could be applied to a number of other cosmological
scenarios where it is important to include momentum density.  It would also be
interesting to compare these methods to those proposed in the recent work of
\cite{Aurrekoetxea:2022mpw}, where the authors construct initial data with
non-trivial matter configurations by using a new scheme based on the Conformal
Transverse-Traceless (CTT) formalism to solve for both constraint equations. In
this so-called CTTK approach, rather than choosing a constant value for the
trace of the extrinsic curvature $K$, one chooses an initial profile for the
conformal factor, and solves an algebraic equation for a spatially varying $K$.

\acknowledgments

We are grateful to Andrei Linde, Leonardo Senatore, Josu Aurrekoetxea, and
Katy Clough for
helpful discussions about inflationary cosmologies.
The authors acknowledge support from an NSERC Discovery grant.
This research was supported in part by
Perimeter Institute for Theoretical Physics.
Research at Perimeter Institute is supported in
part by the Government of Canada through the
Department of Innovation, Science and Economic Development
Canada and by the Province of Ontario through the
Ministry of Economic Development, Job Creation and
Trade. This research was enabled in part by support
provided by SciNet (www.scinethpc.ca) and Digital Research Alliance of Canada 
(alliancecan.ca). Calculations were
performed on the Symmetry cluster at Perimeter Institute, 
the Niagara cluster at the University of
Toronto, and the Narval cluster at Ecole de technologie sup\'erieure
in Montreal.

\appendix
\section{Numerical methodology}\label{app:numerics}

We solve the equations of motion
\eqref{eq:einstein} and \eqref{eq:eom_phi}
using the generalized harmonic formulation, as described
in \cite{Pretorius:2004jg}.
The numerical scheme we use follows that of~\cite{East:2011aa},
which we briefly summarize here.
We discretize the partial differential equations in space, using standard
fourth-order finite difference stencils, and in time, using fourth-order
Runge-Kutta integration.
We control high frequency numerical
noise using Kreiss-Oliger dissipation~\cite{1972Tell...24..199K}.
We use constraint damping to control the constraint
violating modes sourced by truncation error, with
damping parameter values similar to those used 
in black hole formation using the generalized harmonic
formulation~\cite{Pretorius:2004jg}.
We fix the gauge freedom through specifying the source
functions $H^\alpha$, choosing damped harmonic coordinates 
\cite{Choptuik:2009ww,Lindblom:2009tu},
as in \cite{East:2012mb,East:2015ggf}.
During the expansion phase,
we dynamically adjust the time step size in proportion to
the decreasing global minimum of $1/N$ where $N$ is the lapse
(this would be $N=1/a^3$ in a homogeneous FLRW universe with harmonic time)
in order to avoid violating the
Courant-Friedrichs-Lewy condition~\cite{East:2015ggf,East:2017qmk,Corman:2022rqo}.

Following \cite{Pretorius:2004jg}, we track the evolution of any apparent
horizons by finding the surfaces where the outer null expansion vanishes, and
excise an ellipsoid-shaped region interior to the horizon. We typically set the
ratio of the maximum ellipsoid axis to the maximum black hole radial value to
be $0.78$.

The simulation domain is three dimensional with length $L$ in each direction.
The simulations are performed with between 384 and 512 points across each linear dimension.
We construct initial data describing an inhomogeneous cosmology as described in 
section~\ref{sec:matter} and \ref{sec:metric} using the 
the conformal thin sandwich formalism, as described in \cite{East:2012zn}.
The elliptic equations are solved using a second-order accurate multigrid scheme
with a typical resolution of 384 points across each linear dimension.

\begin{figure}[h]
\centering
\includegraphics[width=0.45\textwidth]{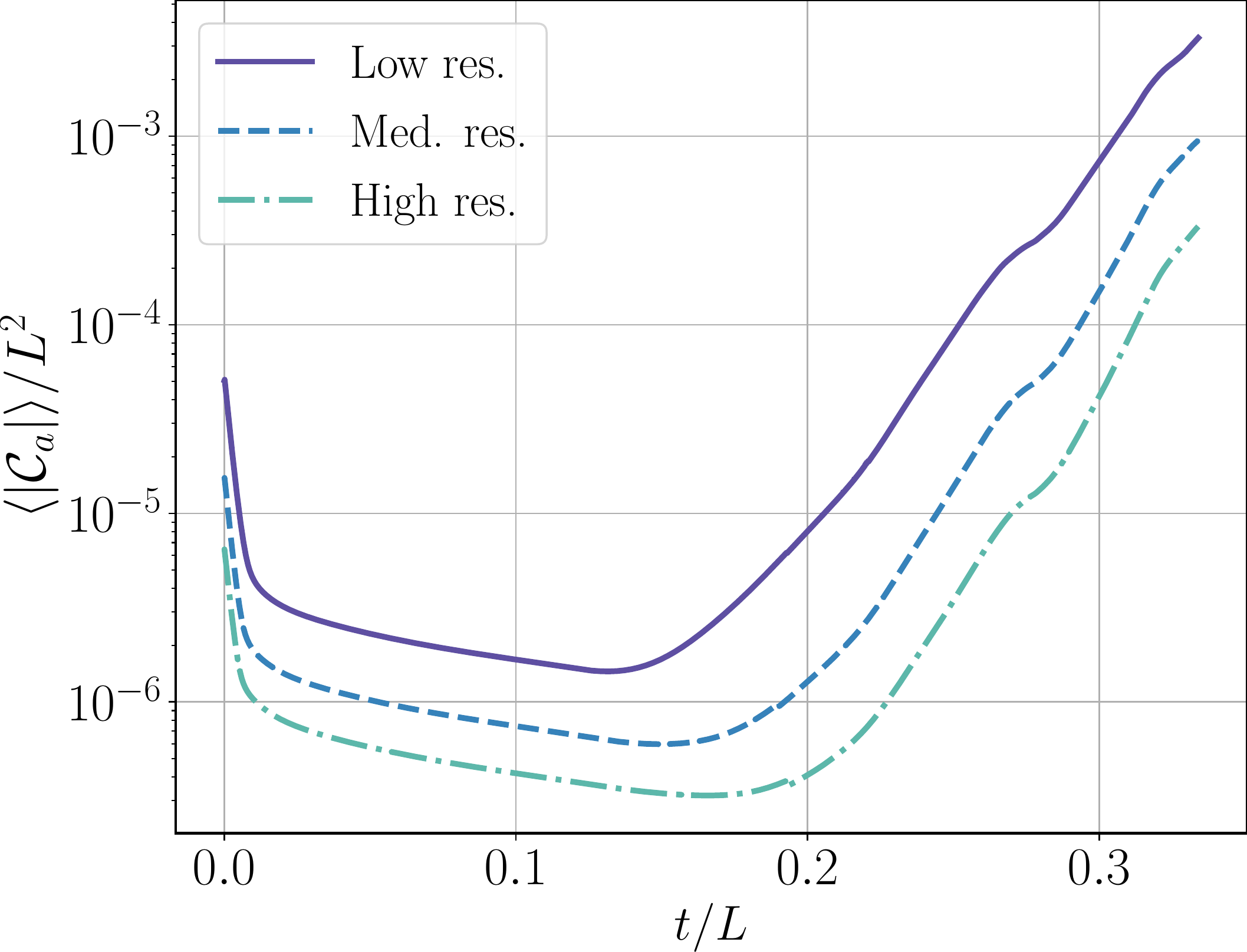}
	\caption{\label{fig:cnv} Volume integrated norm of the constraint violation as
	a function of time for the case shown in right panel of figure~\ref{fig:rho_Aij_NORF}.
	The medium and high resolution have $3/2 \times$ and $2 \times$ the resolution of the low
	resolution. We observe second order convergence at first, followed by fourth order convergence,
	which is consistent with the second order convergence of our initial data code
	and fourth order convergence of our evolution code.
}
\end{figure}

Finally, we present a convergence test of our code and setup.
In figure~\ref{fig:cnv}, we show the time evolution of the norm of 
the constraint violation $C^{\alpha}\equiv \Box x^{\alpha}-H^\alpha$ 
for the strong-field case considered in figure~\ref{fig:rho_Aij_NORF} for different numerical
resolutions.
For this case, the lowest resolution is 256 across each linear dimension
for the evolution and 192 for the initial data code.
The medium and high resolutions correspond, respectively, to an increased
resolution of $3/2$ and $2 \times$ that of the lowest resolution, both in the initial
data and evolution code. We find that the constraints initially converge to zero at roughly
second order and then eventually transition
to fourth order convergence. 
This is because the truncation error is dominated by the second order
convergence of the initial data code at first, but eventually the truncation error
of fourth order evolution code takes over. The high resolution in the convergence study
is equivalent to the resolution we use for most of the cases studied here.

\section{Modified CTS formalism}\label{app:MCTS}
In some cases studied here, instead of solving both the momentum and Hamiltonian constraint using the 
the CTS formalism \cite{East:2012zn}, we construct initial data by 
following a similar approach to \cite{Garfinkle:2008ei,Xue:2013bva},
generalized to three dimensions, 
and choosing initial data such that the momentum constraint is automatically satisfied. 
We then solve for the conformal factor using the Hamiltonian constraint. 
Starting from the constraint equations in the CTS form \eqref{eq:conformal_eqns}, 
choosing conformally flat initial data 
$\tilde{\gamma}_{ij}=\delta_{ij}$, choosing the lapse and shift vector to be 
$N=1$ and $\beta^i=0$, and letting the extrinsic curvature scalar 
be a constant given by \eqref{eq:trK},
the Hamiltonian and momentum constraint equations 
for a stress energy tensor given by \eqref{eq:T} become
\begin{eqnarray}\label{eq:cnsts}
	\partial_i \partial^i\Psi &=&  -\left(\frac{1}{8}\hat{A}_{ij}\hat{A}^{ij}+
	\frac{1}{4}\hat{\eta}^2\right)\Psi^{-7}
	+\left(\frac{1}{12}K^2-\frac{1}{2}V\right)\Psi^5 
	+\frac{1}{4} \partial_i \phi \partial^i \phi \Psi, \nonumber \\
\partial_j \hat{A}^{ij} &=& -2\hat{\eta} \delta^{ij} \partial_j \phi
\end{eqnarray}
where 
\begin{equation}
\hat{\eta} = \Psi^6 \eta= \Psi^6 \partial_t \phi 
\end{equation}
and in last step we assumed $N=1$ and $\beta^i=0$. 
Let us consider the case where we have inhomogeneities along all spatial dimensions. 
Then the momentum constraint is solved by the following ansatz:
\begin{eqnarray}
	\hat{\eta}(x)&=&\frac{1}{\sqrt{2}}\left[{\hat{\eta}}_0 +f_0 \cos (kx)+e_0 \cos (ky) +d_0 \cos (kz)\right], \nonumber \\
	\phi(x) &=& \frac{1}{\sqrt{2}}\left[ \hat{\phi}_0 +f_1 \cos (kx)+ e_1 \cos (ky) +d_1 \cos (kz)\right],
\end{eqnarray}
and the particular solution 
\begin{equation}\label{eq:Aij_mcts}
\hat{A}^{ij}=
\begin{pmatrix}
\hat{A}^{11}(x,y,z) & \hat{A}^{12}(x,y) & \hat{A}^{13}(x,z) \\
\hat{A}^{12}(x,y) & \sigma \hat{A}^{11} (x,y,z) & \hat{A}^{23}(y,z) \\
\hat{A}^{13}(x,z)& \hat{A}^{23}(y,z) & -(1+\sigma) \hat{A}^{11} (x,y,z)
\end{pmatrix} ,
\end{equation}
where $f_0,\ f_1,\ e_0,\ e_1,\ d_0,\ d_1$, and $\sigma$ are parameters to choose and
\begin{eqnarray}
\hat{A}^{11}(x,y,z) &=& -\hat{\eta}_0 f_1 \cos (kx) -\frac{1}{4}f_0 f_1 \cos(2 k x)-\hat{\eta}_0 \frac{e_1}{\sigma} \cos (ky) -\frac{1}{4} \frac{e_0 e_1}{\sigma} \cos(2 k y) \nonumber \\
&& +\hat{\eta}_0 \frac{d_1}{1+\sigma} \cos (kz) +\frac{1}{4} \frac{d_0 d_1}{1+\sigma} \cos(2 k z) \\
\hat{A}^{12}(x,y) &=& e_0 f_1 \sin (kx) \sin (ky) , \qquad {\mathrm{where}} \ \ e_0 f_1 = e_1 f_0 \\
\hat{A}^{13}(x,z) &=& d_0 f_1 \sin (kx) \sin (kz) , \qquad {\mathrm{where}} \ \ d_0 f_1 = d_1 f_0 \\
\hat{A}^{23}(x,z) &=& d_0 e_1 \sin (ky) \sin (kz) , \qquad {\mathrm{where}} \ \ d_0 e_1 = d_1 e_0 
\end{eqnarray}
The conditions on the constants $f_0,\ f_1,\ e_0,\ e_1,\ d_0,$ and $d_1$ 
come from substituting the ansatz into \eqref{eq:cnsts}.

\bibliography{main}
\bibliographystyle{JHEP}

\end{document}